\documentclass[english,12pt]{article}
\usepackage[T1]{fontenc}
\usepackage[utf8]{inputenc}
\usepackage{times}
\usepackage{lmodern}
\usepackage[english]{babel}
\usepackage[pdftex]{graphicx}
\usepackage[width=\textwidth]{caption}
\usepackage[top=1in, bottom=1in, left=1in, right=1in]{geometry}
\usepackage{amsmath,amsopn,amssymb,bm, dsfont,mathrsfs,mathtools}
\usepackage{array,float, placeins,flafter, longtable,booktabs,pdflscape}
\usepackage{verbatim}
\usepackage{color}
\usepackage{natbib}
\usepackage{xr}
\usepackage{pdfpages}
\usepackage[shortlabels]{enumitem}
\usepackage{ulem}

\makeatletter

\newtheorem{thm}{Theorem}

\newtheorem{prop}{Proposition}
\newtheorem{lem}{Lemma}
\newtheorem{conddes}{Design Restriction}
\newtheorem{des}{Design}
\newtheorem{hyp}{Assumption}
\newtheorem{reg}{Regression}

\newtheorem{mydef}{Definition}

\newcommand{\0}{\bm{0}}
\newcommand{\1}{\bm{1}}
\newcommand{\ind}[1]{\mathds{1}\left\{#1\right\}}

\newcommand{\eps}{\varepsilon}

\newcommand{\convP}{\stackrel{P}{\longrightarrow}}
\newcommand{\convL}{\stackrel{d}{\longrightarrow}}

\newcommand{\st}[1]{\texttt{#1}}
\newcommand{\pl}{\text{pl}}

\newcommand{\DID}{\text{DID}}
\newcommand{\DIDM}{\DID_{\text{M}}}
\newcommand{\CI}{\text{CI}}

\newcommand*{\storecounter}[2]{%
  \edef\@currentlabel{\the\value{#1}}
  \label{#2}
}

\date{}

\begin{document}

\title{Difference-in-Differences Estimators of Intertemporal Treatment Effects\thanks{We are very grateful to Diego Ciccia, Felix Knau, M\'{e}litine Mal\'{e}zieux and Doulo Sow for their outstanding research assistance. We are also very grateful to Isabelle M\'ejean, Ashesh Rambachan, Jonathan Roth, Pedro Sant'Anna, Jesse Shapiro, the editor, and four anonymous referees for their helpful comments. Cl\'{e}ment de Chaisemartin was funded by the European Union (ERC, REALLYCREDIBLE,GA N°101043899). Views and opinions expressed are those of the authors and do not reflect those of the European Union or the European Research Council Executive Agency. 
Xavier D’Haultf\oe uille gratefully acknowledges financial support from the research grants Otelo (ANR-17-CE26-0015-041).}}

\author{Cl\'{e}ment de Chaisemartin\thanks{Economics Department, Sciences Po, clement.dechaisemartin@sciencespo.fr%
} \and Xavier D'Haultf\oe{}uille%
\thanks{CREST-ENSAE, xavier.dhaultfoeuille@ensae.fr.}}

\maketitle ~\vspace{-1.2cm}

\begin{abstract}
We study treatment-effect estimation using panel data. The treatment may be non-binary, non-absorbing, and the outcome may be affected by treatment lags. We make a parallel-trends assumption, and propose event-study estimators of the effect of being exposed to a weakly higher treatment dose for $\ell$ periods. We also propose normalized estimators, that estimate a weighted average of the effects of the current treatment and its lags. We also analyze commonly-used two-way-fixed-effects regressions. Unlike our estimators, they can be biased in the presence of heterogeneous treatment effects. A local-projection version of those regressions is biased even with homogeneous effects.
\end{abstract}



\section{Introduction} 
\label{sec:introduction}

We study  treatment-effect estimation, using a panel of groups, indexed by $g$, observed at several time periods, indexed by $t$. $Y_{g,t}$, group $g$'s period-$t$ outcome, may be affected by $D_{g,t}$, group $g$'s period-$t$ treatment, but also by $g$'s lagged treatments.

\medskip
In such instances, to estimate treatment effects, researchers often use two-way fixed effects (TWFE) regressions of $Y_{g,t}$ on group fixed effects, period fixed effects, and some measures of contemporaneous and past exposure to the treatment. A recent literature has shown that TWFE regressions are not robust to heterogeneous treatment effects across groups and/or over time.
Another strand of literature has proposed heterogeneity-robust difference-in-differences (DID) estimators, relying on parallel-trends assumptions, like TWFE estimators, but robust to heterogeneous treatment effects. In that second strand, most of the papers that allow lagged treatments to affect the outcome assume that treatment is binary and staggered, meaning absorbing. In this paper, we propose heterogeneity-robust DID estimators, in applications where the treatment is either non-binary and/or non-absorbing, and where the lagged treatments may affect the outcome. Applications with a non-binary and/or non-absorbing treatment are common. Of the 100 most-cited papers published by the American Economic Review (AER) from 2015 to 2019, we find that 26 estimate a TWFE regression, but only four have a binary-and-absorbing treatment. Our estimators are widely applicable: essentially, they can be used in any design where some groups keep their period-one treatment for a few periods. They are computed by  the \st{did\_multiplegt\_dyn} Stata package \citep{de2023did_multiplegt_dyn}.\footnote{\st{did\_multiplegt\_dyn} supersedes the \st{did\_multiplegt} Stata package, to estimate treatment effects allowing for effects of lagged treatments on the outcome.}

\medskip
Let $Y_{g,t}(d_1,...,d_t)$ denote the potential outcome of $g$ at $t$, if its treatments from period $1$ to $t$ are equal to $(d_1,...,d_t)$. Let $F_g$ denote the first period at which group $g$'s treatment changes, and
$$\delta_{g,\ell}=E(Y_{g,F_g-1+\ell}-Y_{g,F_g-1+\ell}(D_{g,1},...,D_{g,1}))$$
be the expected difference between $g$'s actual outcome at $F_g-1+\ell$ and the counterfactual ``status quo'' outcome it would have obtained if its treatment had remained equal to its period-one value from period one to $F_g-1+\ell$. To estimate $\delta_{g,\ell}$, we propose an estimator $\DID_{g,\ell}$ comparing the $F_g-1$-to-$F_g-1+\ell$ outcome evolution between group $g$, and groups whose treatment has not changed yet at $F_g-1+\ell$, and with the same treatment as $g$ at period one. This last requirement is important: we show that comparing switchers and non-switchers with different period-one treatments relies on a parallel-trends assumption that, essentially, rules out both effects of the lagged treatments on the outcome and time-varying treatment effects. To test the weaker parallel-trends assumption underlying our $\DID_{g,\ell}$ estimator, we propose placebo estimators comparing the outcome trends of switchers and non-switchers with the same period-one treatment, before switchers switch.

\medskip
We then aggregate the $\DID_{g,\ell}$ estimators across groups. We discard from the aggregation the $\DID_{g,\ell}$s such that at $F_g-1+\ell$, $g$ has experienced both a strictly lower and a strictly larger treatment than its period-one treatment: for such $(g,\ell)$s, $\delta_{g,\ell}$ can be written as a linear combination, with negative weights, of the effects of increasing different treatment lags. Thus, $\delta_{g,\ell}$ does not satisfy the following no-sign reversal property: one may have that $(d_1,...,d_t)\mapsto Y_{g,t}(d_1,...,d_t)$ is increasing in each of its arguments, but $\delta_{g,\ell}$ is negative.\footnote{\label{footnote:nosignreversal}The no-sign reversal property was introduced in a static model where the outcome is only affected by the current treatment \citep{small2007stochastic}. When the outcome can be affected by the current and lagged treatments, this property is well-defined only under the assumption that all those treatments affect the outcome in the same direction. Otherwise, an estimand can never be of the ``wrong'' sign. The assumption that the current and lagged treatments all affect the outcome in the same direction is of course very strong. Thus, satisfying our extension of no-sign reversal is an even more minimal requirement than in a static model.}  Among the remaining $(g,\ell)$s, either
\begin{align*}
& D_{g,F_g}> D_{g,1},D_{g,F_g+1}\geq D_{g,1},...,D_{g,F_g-1+\ell}\geq D_{g,1},\\
\text{or }&D_{g,F_g}<D_{g,1},D_{g,F_g+1}\leq D_{g,1},...,D_{g,F_g-1+\ell}\leq D_{g,1}.
\end{align*}
In the former case, $\delta_{g,\ell}$ is the effect of having been exposed to a weakly higher treatment for $\ell$ periods, but in the latter case $\delta_{g,\ell}$ is the effect of having been exposed to a weakly lower treatment for $\ell$ periods. Accordingly, when we aggregate the $\DID_{g,\ell}$ estimators, we multiply by minus one the $\DID_{g,\ell}$ for groups such that $D_{g,F_g}<D_{g,1}$. This finally yields a $\DID_{\ell}$ estimator of the effect of having been exposed to a weakly higher treatment dose for $\ell$ periods.

\medskip
In general, lending a more ``structural'' interpretation to the $\DID_{\ell}$ estimators—beyond their reduced-form interpretation—can be challenging as our estimation
method accommodates a range of potentially complicated designs. For instance, with three periods and three groups such that $(D_{1,1}=0,D_{1,2}=4,D_{1,3}=0)$, $(D_{2,1}=0,D_{2,2}=2,D_{2,3}=3)$, and $(D_{3,1}=0,D_{3,2}=0,D_{3,3}=0)$, $\DID_{2}$ estimates the average of $E(Y_{1,3}(0,4,0)-Y_{1,3}(0,0,0))$ and $E(Y_{2,3}(0,2,3)-Y_{2,3}(0,0,0))$, so $\DID_{2}$ does not estimate by how much the outcome increases on average when the treatment increases by a given amount for a given number of periods.

\medskip
In light of these challenges, we propose three strategies for interpretable estimates that may be of interest to researchers. First, when the number of $F_g-1$-to-$F_g-1+\ell$ treatment trajectories is low relative to the number of groups, one may estimate trajectory-specific versions of $\DID_{\ell}$, which may be more easily interpretable. Second, we propose to normalize our $\DID_{\ell}$ estimators by the average of $\sum_{k=1}^\ell|D_{g,F_g-1+k}-D_{g,1}|$, the total treatment increments from $F_g$ to $F_g-1+\ell$ that generate $\delta_{g,\ell}$. We show that $\DID^n_{\ell}$, the normalized $\DID_{\ell}$, is unbiased for a weighted average of the effect of the current treatment and of its $\ell-1$ first lags on the outcome. We also show that
$\DID^n_{\ell}$ can be used to test a null hypothesis which is often of great interest, namely whether the current and lagged treatments all have the same effect on the outcome. Third, we show that a weighted sum across $\ell$ of the $\DID_{\ell}$s is unbiased for a parameter with a clear economic interpretation, which may be used to conduct a cost-benefit analysis assessing if the treatment changes that took place over the panel led to a better situation than the one that would have prevailed if no change had been undertaken, a natural policy question. Importantly,
that parameter can also be interpreted as an average total effect per unit of treatment, where ``total effect'' refers to the sum of the effects of a treatment increment, at the time when it takes place and at later periods.

\medskip
Finally, we derive the asymptotic distribution of our estimators, and confidence intervals for our target parameters. Consistent with the spirit of our paper, those results are derived under weak conditions on the design.

\medskip
In our aforementioned census of highly-cited TWFE papers published in the AER, papers that do not have a binary-and-staggered treatment and estimate dynamic effects do so using three different methods. In some papers, there is no variation in treatment timing: all treated groups start getting treated at the same date, with group-specific treatment intensities. Then, researchers have estimated TWFE regressions of the outcome on the treatment intensity interacted with period fixed effects. In more complicated designs where there may be variation in treatment timing and a group's treatment may increase or decrease multiple times, some researchers have estimated a panel-data version of the local-projection method proposed by \cite{jorda2005estimation}, while other researchers have estimated TWFE regressions of the outcome on the treatment and its first $K$ lags, the so-called distributed-lag regression. We show that under the parallel-trends assumption underlying our estimators, those three regressions may not estimate a convex combination of the $\delta_{g,\ell}$s. Thus, they may not satisfy the no-sign reversal property. The local-projection and distributed-lag regressions also suffer from a contamination problem: coefficients supposed to estimate the effect of a given length of exposure to treatment or of a given treatment lag are actually contaminated by effects of other exposure lengths or other lags. Strikingly, the local-projection regression may yield biased estimators, possibly of the wrong sign, even if the treatment effect is constant across groups and over time.
Therefore, despite their limitations, our estimators improve on current practice.

\medskip
We use our estimators to revisit \cite{favara2015credit}. These authors use the differential timing and intensity of banking deregulations in US states during the 1990s to estimate deregulations' effects on credit supply. Our estimators suggest that banking deregulations have very persistent effects on the growth of credit supply and house prices. These findings differ sharply from those obtained by \cite{favara2015credit}. Using the panel-data version of the local-projection method, the authors find significant short-run effects, which quickly vanish. Strikingly, our decompositions show that local-projection estimators of the long-run effects of deregulations estimate weighted sums of effects, where the sum of the weights is negative. Thus, even if treatment effects are constant, those coefficients are of a different sign than the effect.

\medskip
The paper is organized as follows. Section 2 introduces the set up and our assumptions. Section 3 defines our parameters of interest and presents our estimators. Section 4 studies the properties of commonly used TWFE regressions to estimate current and lagged treatment effects in general designs. Section \ref{sec:appli} uses our estimators to revisit \cite{favara2015credit}. Finally, the Web Appendix details several extensions and gathers all the proofs.

\subsection*{Related literature} 
\label{sec:litt}

Our paper extends previous literature on heterogeneity-robust DID estimators, when lagged treatments can affect the outcome \citep[see, e.g.,][]{callaway2018,abraham2018,borusyak2020revisiting}, because prior papers assume that treatment is binary and absorbing, unlike this paper. Yet, there are cases, reviewed in details at the end of Section \ref{sub:AvsSQunnorm}, where after some relabelling, our non-normalized estimators are numerically equivalent to some estimators of \cite{callaway2018}. Thus, the main contributions we make to that strand of literature amount to highlighting $\DID_\ell$'s lack of ``structural'' interpretation in the complex designs we consider, and proposing three strategies to obtain more interpretable estimators.

\medskip
This paper is also related to our prior work in \cite{dCdH_AER}. There, we also considered potentially non-binary and/or non-absorbing treatments, but we assumed that lagged treatments do not affect the outcome. There are two differences between this paper and our prior work. First, when lagged treatments can affect the outcome and the treatment is non-binary and/or non-absorbing, the $\DIDM$ estimator in \cite{dCdH_AER} may not unbiasedly estimate the effect of the contemporaneous treatment on the outcome, unlike the $\DID_1$ estimator in this paper.\footnote{Consider a binary treatment that may turn on an off. Assume that group $g$ is untreated at period 1, untreated at period 2, and treated at period 3. Ruling out dynamic effects, any group $g'$ that is untreated at periods 2 and 3 is a valid control group under a parallel trends assumption on $Y_{g,t}(0)$. However, in that set of control groups, there may be some groups that were treated at period $1$. Allowing for dynamic effects, such $g'$ become invalid control groups: their period-one treatment may still affect their period-two-to-three outcome evolution. Therefore, unlike $\DIDM$, the $\DID_1$ estimator in this paper does not use those $g'$ as control groups.} Second, with a non-binary treatment, \cite{dCdH_AER} also consider a normalized estimator, but the normalization in this paper is different. \cite{dCdH_AER} normalize the effect of switches by switchers' average treatment change, to ensure $\DIDM$ has a slope interpretation. Here, because potential outcomes are a function of several arguments (the current and the lagged treatments), the normalization that leads to the ``weighted average of the effects of different treatment lags'' interpretation divides $\DID_\ell$ by the sum of switchers' treatment increments from $F_g$ to $F_g+\ell$.

\medskip
In work posterior to ours, \cite{callaway2021difference} consider designs where groups are all untreated at period one, and then get treated at heterogeneous dates, with heterogeneous intensities. Our estimators can also be used in those designs. Some of the estimands they propose compare the outcome evolution of treated and not-yet-treated, like ours.

\medskip
An important literature has proposed estimators of the effects of current and lagged treatments, when the treatment can both increase or decrease over time, under a sequential ignorability assumption. This assumption requires that at each period, the treatment is independent of current or future potential outcomes, conditional on past treatments and outcomes \citep[see][]{robins1986new,Murphy2001,bojinov2020panel}. Instead, we consider a parallel-trends assumption, and consider alternative estimands and estimators under that assumption.

\medskip
Finally, our decompositions of the regression coefficients used by applied researchers to estimate dynamic effects in general designs are related to the literature showing that TWFE regressions are not robust to heterogeneous treatment effects \citep[see, e.g.,][]{dCdH_AER,goodman2018,borusyak2016}, and in particular to \cite{abraham2018} and \cite{dCDH_several}, who respectively study event-study regressions with a binary and staggered treatment, and regressions with several treatments. Our main contribution to this literature is our decomposition of the TWFE local-projection regressions. The fact that those regressions are misspecified and could exhibit sign reversal even under constant effects is unique to that particular class of TWFE regressions.

\section{Setup, design, and identifying assumptions}\label{sec:the_set_up}

\subsection{Setup}

\paragraph{Group-level panel data.} We have a panel of $G$ groups observed at $T$ periods, respectively indexed by $g$ and $t$. Typically, groups are geographical entities, like states or counties, but a group could also just be a single individual or firm. The group-level panel data may be constructed by aggregating an individual-level panel or repeated cross-section data set at the $(g,t)$ level, defining groups, say, as individuals' county of birth. The group-level panel data may also be constructed from a single cross-section, with cohort of birth playing the role of time.
The estimators we propose below are not weighted by $N_{g,t}$, the population of cell $(g,t)$. This is just to reduce notational complexity: proposing weighted estimators is a mechanical extension.\footnote{The \st{did\_multiplegt\_dyn} command that implements the estimators proposed in this paper can be used with data at a more disaggregated level than the $(g,t)$ level. Then, it aggregates the data at the $(g,t)$ level internally and automatically weights $(g,t)$ cells by their number of observations in the data. To use different weights, or to weight the estimation when working directly with a group-level panel data, one can use the \st{weight} option.}

\paragraph{Treatment.} Let $D_{g,t}$ denote the treatment of group $g$ at period $t$.
Throughout the paper, we assume that the treatment is non-negative: $D_{g,t}\geq 0$, as is most often the case in applications.\footnote{When $D_{g,t}\geq \underline{d}<0$, one can redefine the treatment as $D_{g,t}-\underline{d}$, a non-negative treatment.} Let $\bm{D}_g=(D_{g,1},...,D_{g,T})$ be a vector stacking $g$'s treatments from period $1$ to $T$, and let $\bm{D}=(\bm{D}_1,...,\bm{D}_G)$ be a vector stacking the treatments of all groups at every period. We refer to $\bm{D}$ as the study's design. Finally, let $\mathcal{D}$ be the set of values $\bm{D}_g$ can take (i.e.: its support).

\paragraph{Potential outcomes.} For all $(d_1,...,d_T)\in \mathcal{D}$, let $Y_{g,t}(d_1,...,d_T)$ denote the potential outcome of group $g$ at $t$ if $(D_{g,1},...,D_{g,T})=(d_1,...,d_T)$, and let $Y_{g,t}=Y_{g,t}(\bm{D}_g)$ denote the observed outcome of $g$ at $t$. This dynamic potential outcome framework follows \cite{robins1986new}. It explicitly allows groups' outcome at $t$ to depend on their past and future treatments.

\paragraph{Initial conditions.}
Some observations may have already been treated prior to period 1, and those treatments may still affect some of their period-$1$-to-$T$ outcomes, the so-called initial-conditions problem. However, we cannot estimate those effects, as treatments and outcomes are not observed before period 1, so we do not account for this potential dependency in our notation. We discuss strategies to account for this dependency in Section \ref{sub:lagged} of the Web Appendix.

\paragraph{Conditioning on the design.} When defining our target parameters, we take the perspective of a social planner, seeking to conduct a cost-benefit analysis comparing groups' actual treatments $\bm{D}$ to the counterfactual ``status-quo'' scenario where every group would have kept all the time
the same treatment as in period 1. Thus, the planner wants to know if the treatment/policy changes that took place over the duration of the panel led to a better situation than the one that would have prevailed without any policy change, a natural policy question. Then, our analysis is conditional on $\bm{D}$, the study's design. This implies that our parameters of interest are dictated by the design, rather than chosen by the researcher. As such, they do not exhaust the set of all possible treatment effect parameters one may be interested in, and only partially uncover the structural function mapping the current, lagged, and future treatments to the average outcome. Conditional on the design, only groups' potential outcomes are random, and probabilistic statements below are with respect to their joint probability distribution. We take a model-based approach to uncertainty: potential outcomes are random because of shocks.

\subsection{Design}

\paragraph{Estimators applicable to staggered first switch designs.} The date at which a group's treatment changes for the first time plays a key role in our analysis.

\begin{mydef}(First treatment change) For all $g$, let $F_g=\min\{t:t\geq 2,D_{g,t}\ne D_{g,t-1}\}$.
\end{mydef}\label{def_Fg}
We adopt the convention that $F_g=T+1$ if $g$'s treatment never changes. Our estimators are applicable to any design that satisfies Restriction \ref{hyp:non_pathological_design} below, hereafter referred to as ``staggered first switch designs'', meaning that groups experience their first treatment change at different points in time.
\begin{conddes}\label{hyp:non_pathological_design}
$\exists (g,g')$ such that: (i) $D_{g,1}=D_{g',1}$, (ii) $F_g\ne F_{g'}$.
\end{conddes}
(i) requires that there exist groups with the same period-one treatment. This is essentially a restriction on the support of the period-one treatment. If groups' period-one treatments are i.i.d. draws from a continuous distribution, $D_{g,1}\ne D_{g',1}$ for all $(g,g')$, so (i) fails. On the other hand, if $D_{g,1}$ can only take a finite number of values $K$, (i) automatically holds as long as $K<G$. In Section \ref{sub:continuous} of our Web Appendix, we extend our estimators to designs where (i) fails, but for expositional purposes we prefer to assume throughout the paper that (i) holds. (ii) requires that there is heterogeneity in the date at which groups change treatment for the first time. (ii) fails if groups' treatment is extremely non-persistent, so that $D_{g,1}\ne D_{g,2}$ and $F_g=2$ for all $g$. (ii) also fails if all groups keep their period-one treatment throughout the panel: $F_g=T+1$ for all $g$, or if groups all change treatment for the first time at the same date, for instance due to a universal policy affecting them all: $F_g=t_0$ for all $g$. Overall, though we impose (i) in the paper to simplify exposition, the only designs where our estimators are inapplicable are those where groups' treatment is extremely non-persistent, and designs with a universal treatment change.

\paragraph{Some commonly-found designs where Design Restriction \ref{hyp:non_pathological_design} often holds.}
\begin{des}\label{des:binaryandstaggered}
(Binary and staggered treatment) $\forall$ $(g,t)$, $D_{g,t}=1\{t\geq F_g\}$, with $F_g \geq 2$.
\end{des}
Design \ref{des:binaryandstaggered} has been studied by \cite{callaway2018}, \cite{goodman2018}, \cite{abraham2018}, and \cite{borusyak2020revisiting}, among others. Assuming that all groups are initially untreated ($F_g \geq 2$) is without loss of generality: in Design \ref{des:binaryandstaggered}, the treatment effects of always-treated groups cannot be estimated under parallel-trends assumptions.
\begin{des}\label{des:binaryoneexit}
(Binary treatment, groups can join and then leave treatment) $\forall$ $(g,t)$, $D_{g,t}=1\{E_g\geq t\geq F_g\}$, with $2\leq F_g \leq E_g$.
\end{des}
In Design \ref{des:binaryoneexit}, groups may get treated and leave treatment. For instance, the design of \cite{burgess2015value} is very close to Design \ref{des:binaryoneexit}.\footnote{Specifically, for the co-ethnicity $\times$ democracry treatment in their Equation (2), districts can enter and leave treatment at most twice over their entire study period.} An important special case of Design \ref{des:binaryoneexit} is when treated groups are treated for only one period ($E_g=F_g$), leading to
\begin{equation}\label{eq:oneshottreatment}
D_{g,t}=1\{t=F_g\},
\end{equation}
which we hereafter refer to as a ``one-shot-treatment design''.

\begin{des}\label{des:stag_nonbinary}
(Staggered design with group-specific intensities) $\forall$ $(g,t)$, $D_{g,t}=I_g1\{t\geq F_g\}$, $F_g \geq 2$.
\end{des}
In Design \ref{des:stag_nonbinary}, there is variation across groups in treatment timing and/or intensity. For instance, the design of \cite{pierce2016surprisingly} corresponds to Design \ref{des:stag_nonbinary}.
\begin{des}\label{des:untreatedatbaseline}
(Zero treatment at baseline) $\forall$ $g$, $D_{g,1}=0$.
\end{des}
Designs \ref{des:binaryandstaggered}-\ref{des:stag_nonbinary} are special cases of Design \ref{des:untreatedatbaseline}. We single out Design \ref{des:untreatedatbaseline}, because such designs are frequent. Our estimators can be used outside of that design, like in Design \ref{des:discrete} below.
\begin{des}\label{des:discrete}
(Discrete treatment at baseline) $\forall$ $g$, $D_{g,1}\in \{0,1,...,K\}$.
\end{des}
In Design \ref{des:discrete}, groups' baseline treatment is discrete, and treatment paths are unrestricted. For instance, the design of \cite{fuest2018higher}  corresponds to Design \ref{des:discrete}. 
A special case of Design \ref{des:discrete} is when groups' baseline treatment is binary.

\subsection{Identifying assumptions}

\paragraph{No-anticipation.} We maintain throughout the paper the following condition.

\begin{hyp}\label{hyp:no_antic}
	(No Anticipation) $\forall g$, $\forall(d_1,...,d_T)\in \mathcal{D}$, $Y_{g,t}(d_1,...,d_T)=Y_{g,t}(d_1,...,d_t)$.
	\end{hyp}
Assumption \ref{hyp:no_antic} requires that a group's current outcome does not depend on its future treatments, the so-called no-anticipation hypothesis. This assumption appears in \cite{robins1986new}, \cite{abbring2003nonparametric}, \cite{malani2015}, \cite{botosaru2018difference}, \cite{callaway2018}, and \cite{abraham2018}.

\paragraph{Parallel trends.}

Let
$\mathcal{D}^{\text{r}}_1=\{d:\exists (g,g')\in\{1,...,G\}^2:D_{g,1}=D_{g',1}=d,F_g\ne F_{g'}\}$
be the set of period-one-treatment values such that two groups with different values of $F_g$ have that period-one treatment. For any $d$ in $\mathcal{D}^{\text{r}}_1$ and any $t$, let $\bm{d}_t$ denote a $1\times t$ vector of $d$s. For any integer $k$, let $\bm{D}_{g,1,k}$ be a $1\times k$ vector whose coordinates are all equal to $D_{g,1}$. $Y_{g,t}(\bm{D}_{g,1,t})$ is $g$'s period-$t$ outcome in a counterfactual where it keeps its period-one treatment till period $t$. We refer to it as its ``status-quo'' potential outcome.
\begin{hyp}\label{hyp:strong_exogeneity}
	(Parallel trends for the status-quo outcome, for groups with the same period-one treatment)
$\forall (g, g’)$, if $D_{g,1} = D_{g’,1} \in \mathcal{D}_1^{\text{r}}$, then $\forall t\geq 2$,
$$E[Y_{g,t}(\bm{D}_{g,1,t}) - Y_{g,t-1}(\bm{D}_{g,1,t-1}) | \bm{D}] = E[Y_{g’,t}(\bm{D}_{g’,1,t}) - Y_{g’,t-1}(\bm{D}_{g’,1,t-1}) | \bm{D}].$$
\end{hyp}
Assumption \ref{hyp:strong_exogeneity} requires that if two groups have the same period-one treatment, then they have the same expected evolution of their status-quo outcome.  Assumption \ref{hyp:strong_exogeneity} is a generalization of the standard parallel-trends assumption in DID models \citep[see, e.g.,][]{Abadie05} to our set-up that allows for dynamic effects and for potentially complicated designs where groups may not all be untreated at period one. In Design \ref{des:untreatedatbaseline}, under Restriction \ref{hyp:non_pathological_design},
$\mathcal{D}^{\text{r}}_1=\{0\}$, so Assumption \ref{hyp:strong_exogeneity} only restricts the never-treated potential outcome $Y_{g,t}(\bm{0}_t)$, and is similar to the identifying assumption considered by \cite{callaway2018} and \cite{abraham2018} for binary and staggered designs. Note that Assumption \ref{hyp:strong_exogeneity} restricts only one potential outcome per group, so even when $\mathcal{D}^{\text{r}}_1$ contains more than one value, Assumption \ref{hyp:strong_exogeneity} alone does not restrict groups' treatment effects.
If Assumption \ref{hyp:strong_exogeneity} is implausible for some values in $\mathcal{D}^{\text{r}}_1$, the estimators we propose can still be used, excluding groups with $D_{g,1}$ equal to one of those values. Assumption \ref{hyp:strong_exogeneity} has testable implications that are easy to assess, using placebo estimators mimicking the actual estimators we propose (see Section \ref{sec:placebos}).

\section{Target parameters and estimators}
\label{sec:target}

\subsection{Non-normalized actual-versus-status-quo and event-study effects}\label{sub:AvsSQunnorm}

\paragraph{Definition of the non-normalized actual-versus-status-quo effects.}
For every $g$, let
$$T_g=\max_{g':D_{g',1}=D_{g,1}}F_{g'}-1$$
denote the last period where there is still a group with the same period-one treatment as $g$ and whose treatment has not changed since the start of the panel. For any $g$ such that $F_g\leq T_g$, and for any $\ell \in \{1,...,T_g-F_g+1\}$, let
$$\delta_{g,\ell}=E\left[Y_{g,F_g-1+\ell}-Y_{g,F_g-1+\ell}(D_{g,1},...,D_{g,1})|\bm{D}\right]$$
be the expected difference between group $g$'s actual outcome at $F_g-1+\ell$ and its counterfactual ``status quo'' outcome if its treatment had remained equal to its period-one value from period one to $F_g-1+\ell$. We refer to $\delta_{g,\ell}$ as the actual-versus-status-quo (AVSQ) effect of $g$ at $F_g-1+\ell$.

\paragraph{Estimation of the non-normalized actual-versus-status-quo effects.}
For any finite set $A$, let $\#A$ be the number of elements of $A$. For all $(g,t)$, let
$$N^g_t=\#\{g':D_{g',1}=D_{g,1},F_{g'} >t\}$$
be the number of groups $g'$ with the same period-one treatment as $g$, and that have kept the same treatment from period $1$ to $t$. For every $g$ such that $F_g\leq T_g$, and every $\ell\in \{1,...,T_g-F_g+1\}$, $N^{g}_{F_g-1+\ell}>0$. To estimate $\delta_{g,\ell}$, we use
$$\DID_{g,\ell} =Y_{g,F_g-1+\ell} - Y_{g,F_g-1} -
\frac{1}{N^{g}_{F_g-1+\ell}}\sum_{g':D_{g',1}=D_{g,1},F_{g'} >F_g-1+\ell}(Y_{g',F_g-1+\ell} - Y_{g',F_g-1}),$$
a DID estimator comparing the $F_g - 1$-to-$F_g-1+\ell$ outcome evolution of $g$ to that of groups with the same baseline treatment, and that have kept that treatment from period $1$ to $F_g-1+\ell$.\footnote{\label{footnote:efficientestimator} $\DID_{g,\ell}$ uses groups' $F_g - 1$ outcome as the baseline. Instead, one could average their period-1-to-$F_g - 1$ outcomes. This would give rise to another unbiased estimator of $\delta_{g,\ell}$. Which estimator is more precise depends on the outcome's serial correlation \citep[see][]{borusyak2020revisiting,harmon2022difference}. Moreover, this second estimator could be more biased than $\DID_{g,\ell}$ if Assumption \ref{hyp:strong_exogeneity} does not exactly hold and the discrepancy between groups' trends gets larger over longer horizons. See \cite{de2022survey} for further discussion.}
\begin{lem}\label{lem:id_delta_g_ell}
If Assumptions \ref{hyp:no_antic} and \ref{hyp:strong_exogeneity} hold, then for every $(g,\ell)$ such that $1 \le \ell \le T_g-F_g+1$,  $E\left[\DID_{g,\ell}\middle| \bm{D}\right]= \delta_{g,\ell}$.
\end{lem}
\paragraph{No-crossing condition.}
Assume that $g_0$ is such that $D_{g_0,1}=1,D_{g_0,2}=2,D_{g_0,3}=0$. Then,
\begin{align*}
\delta_{g_0,2}=& E[Y_{g_0,3}(1,2,0)-Y_{g_0,3}(1,1,1)] \\ =& E[Y_{g_0,3}(1,2,0)-Y_{g_0,3}(1,1,0)]-E[Y_{g_0,3}(1,1,1)-Y_{g_0,3}(1,1,0)]	
\end{align*}
is the difference between the effect of increasing $g_0$'s period-$2$ treatment from $1$ to $2$, and the effect of increasing $g_0$'s period-$3$ treatment from $0$ to $1$. One could have that both effects are positive but $\delta_{g_0,2}$ is negative. Therefore, $\delta_{g_0,2}$ does not satisfy the no-sign reversal property: $(d_1,d_2,d_3)\mapsto Y_{g,t}(d_1,d_2,d_3)$ may be increasing in each of its arguments but $\delta_{g_0,2}$ is negative. Beyond this example, for all $(g,\ell)$  such that $\ell$ periods after its first treatment change, $g$ has experienced both a treatment strictly below and a treatment strictly above its period-one treatment, $\delta_{g,\ell}$ can be written as a linear combination, with negative weights, of the effects of increasing different treatment lags. Throughout this section, we assume away the existence of such $(g,\ell)$s.
\begin{conddes}\label{hyp:monotonic_design}
$\forall g\in \{1,...,G\}$, either $D_{g,t}\geq D_{g,1}$ $\forall t$, or $D_{g,t}\leq D_{g,1}$ $\forall t$.
\end{conddes}
Restriction \ref{hyp:monotonic_design} is implied by Design \ref{des:untreatedatbaseline}. It also holds automatically when the treatment is binary, or when groups' treatment can only change once. When Restriction \ref{hyp:monotonic_design} fails, one can discard from the sample all cells $(g,t)$ such that at $t$, $g$ has experienced both a treatment strictly below and a treatment strictly above its period-one treatment.\footnote{\st{did\_multiplegt\_dyn} automatically drops those cells if the \st{drop\_larger\_lower} option is specified.} This yields an unbalanced panel of groups where Restriction \ref{hyp:monotonic_design} holds by construction, on which our estimators can be applied. We impose Restriction \ref{hyp:monotonic_design} to avoid the notational burden of defining our estimators on an unbalanced panel.\footnote{\st{did\_multiplegt\_dyn}  can be used with an unbalanced panel of groups, see the help file.}

\paragraph{Definition of the non-normalized event-study effects.}
Let $L=\max_{g}(T_g-F_g+1)$ denote the largest $\ell$ such that $\delta_{g,\ell}$ can be estimated for at least one $g$.
Under Restriction \ref{hyp:non_pathological_design}, $L\geq 1$.
For every $\ell \in \{1,...,L\}$, let
$$N_\ell=\#\{g:F_g-1+\ell\leq T_g\}$$
be the number of groups for which $\delta_{g,\ell}$ can be estimated. For all $g$ such that $F_g\leq T$, let
$$S_g=1\{D_{g,F_g}>D_{g,1}\}-1\{D_{g,F_g}<D_{g,1}\}$$	
be equal to 1 (resp. -1) for groups whose treatment increases (resp. decreases) at $F_g$. Then, let
\begin{align}
\delta_{\ell} =&\frac{1}{N_\ell}\sum_{g:F_g-1+\ell\leq T_g}S_g\delta_{g,\ell}, \label{eq:def_deltaell}
\end{align}
be the average of $S_g\delta_{g,\ell}$. Under Restriction \ref{hyp:monotonic_design}, for groups with $S_g=1$, $D_{g,t}\geq D_{g,1}$ for all $t$, so $\delta_{g,\ell}$ is the effect of having been exposed to a weakly higher treatment dose for $\ell$ periods. Conversely,
for groups with $S_g=-1$, $D_{g,t}\leq D_{g,1}$ for all $t$, so $\delta_{g,\ell}$ is the effect of having been exposed to a weakly lower dose for $\ell$ periods. Taking the negative of $\delta_{g,\ell}$ for those groups ensures that $\delta_{\ell}$ is an average effect of having been exposed to a weakly larger dose for $\ell$ periods.

\paragraph{Estimation of the non-normalized event-study effects.} For every $\ell \in \{1,...,L\}$, let
$$\DID_{\ell} =\frac{1}{N_\ell}\sum_{g:F_g-1+\ell\leq T_g}S_g\DID_{g,\ell}.$$
Lemma \ref{lem:id_delta_g_ell} implies that $\DID_{\ell}$ is conditionally unbiased for $\delta_{\ell}$ under Assumptions  \ref{hyp:no_antic} and \ref{hyp:strong_exogeneity}.

\paragraph{Interpreting non-normalized event-study effects.}
In Design \ref{des:binaryandstaggered}, $$\delta_{\ell}=\frac{1}{N_\ell}\sum_{g:F_g-1+\ell\leq T_g}E\left[Y_{g,F_g-1+\ell}(\0_{F_g-1},\1_{\ell})-Y_{g,F_g-1+\ell}(\0_{F_g-1+\ell})
|\bm{D}\right].$$
Thus, $\delta_\ell$ is just the average effect of having been treated rather than untreated for $\ell$ periods, across all groups reaching $\ell$ treatment periods at or before the last period where there is still at least one untreated group. Outside of Design \ref{des:binaryandstaggered}, the interpretation of $\delta_\ell$ is more complicated. For instance,
even in Design \ref{des:binaryoneexit} ($D_{g,t}=1\{E_g\geq t\geq F_g\}$), for all $g$ such that $F_g-1+\ell> E_g$, group $g$ has exited the treatment at $F_g-1+\ell$, and $$\delta_{g,\ell}=E\left[Y_{g,F_g-1+\ell}(\0_{F_g-1},\1_{E_g-F_g+1},\0_{F_g-1+\ell-E_g})-Y_{g,F_g-1+\ell}(\0_{F_g-1+\ell})
|\bm{D}\right]$$
is the effect of having been treated for $E_g-F_g+1$ periods, $F_g-1+\ell-E_g$ periods before the outcome is measured. Thus, the number and the recency of the treatment periods that generate $\delta_{g,\ell}$ vary across groups, complicating the interpretation of $\delta_{\ell}$. Similarly, with three periods and three groups such that $(D_{1,1}=0,D_{1,2}=4,D_{1,3}=0)$, $(D_{2,1}=0,D_{2,2}=2,D_{2,3}=3)$, and $(D_{2,1}=0,D_{2,2}=0,D_{2,3}=0)$, $\DID_{2}$ estimates the average of $E(Y_{1,3}(0,4,0)-Y_{1,3}(0,0,0))$ and $E(Y_{2,3}(0,2,3)-Y_{2,3}(0,0,0))$. Thus, the magnitude and timing of the treatment increments generating $\delta_{g,\ell}$  varies across groups, which again complicates the interpretation of $\delta_{\ell}$. Still, in any design satisfying Restriction \ref{hyp:monotonic_design}, $\delta_{\ell}$ can be interpreted as an average effect of having been exposed to a weakly higher treatment for $\ell$ periods.

\paragraph{More disaggregated effects.} If the design is such that $P_\ell:=\#\{(D_{g,1},D_{g,F_g},...,D_{g,F_g-1+\ell}):F_g-1+\ell\leq T_g\}$ is very low relative to $G$, the number of period-$1$-to-$F_g-1+\ell$ treatment paths among the groups entering in $\delta_{\ell}$ may be sufficiently low to precisely estimate the average of $\delta_{g,\ell}$ separately across all groups with the same path, thus yielding estimates of the average effects of specific treatment paths.\footnote{For any treatment path $(D_{g,1},D_{g,F_g},...,D_{g,F_g-1+\ell})=\bm{d}$, a treatment-path-specific version of $\delta_{\ell}$ can easily be estimated using the \st{did\_multiplegt\_dyn} command, restricting the sample to $(g,t)$s such that $t<F_g$, or $F_g\leq t\leq F_g-1+\ell$ and $(D_{g,1},D_{g,F_g},...,D_{g,F_g-1+\ell})=\bm{d}$.} For instance, in Design \ref{des:binaryoneexit} the number of treatment paths may often be low enough for this solution to be practical. But in more complicated designs where treatment paths are very group-specific, $P_\ell$ may often be too large for this solution to be practical, especially as $\ell$ increases. In such instances, we still recommend that practitioners report the values in $\{(D_{g,1},D_{g,F_g},...,D_{g,F_g-1+\ell}):F_g-1+\ell\leq T_g\}$, and their distribution: this information may be helpful to interpret $\DID_\ell$.

\paragraph{Comparison with unconditional event-study estimators of \cite{callaway2018} in Design \ref{des:untreatedatbaseline}.}
When all groups have the same period-one treatment, $\DID_\ell$ is equivalent to the estimator obtained by redefining the treatment as an indicator equal to one if group $g$'s treatment has ever changed at $t$, and then computing the estimator of $\theta_{es}(\ell-1)$ of \cite{callaway2018} with this binarized and staggerized treatment.\footnote{This ``binarize and staggerize'' idea has for instance been used by \cite{deryugina2017fiscal} or \cite{krolikowski2018choosing}.}

\paragraph{Comparison with unconditional event-study estimators of \cite{callaway2018} outside of Design \ref{des:untreatedatbaseline}.}
When groups' period-one treatment varies, the two estimators are not equivalent:\footnote{\label{footnote:East} There are recent papers that binarize and staggerize the treatment and compute the unconditional event-study estimators of \cite{callaway2018} outside of Design \ref{des:untreatedatbaseline}, see, e.g. \cite{east2023multi}.}
$\DID_\ell$ only compares switchers and non-switchers with the same period-one treatment, whereas the aforementioned estimator compares switchers and non-switchers with different period-one treatments. Then, that estimator relies on Assumption \ref{hyp:strong_exogeneity2}:
\begin{hyp}\label{hyp:strong_exogeneity2}
	(Parallel trends for the status-quo outcome, for all groups)
$\forall (g, g’)$, $\forall t\geq 2$,
$$E[Y_{g,t}(\bm{D}_{g,1,t}) - Y_{g,t-1}(\bm{D}_{g,1,t-1}) | \bm{D}] = E[Y_{g’,t}(\bm{D}_{g’,1,t}) - Y_{g’,t-1}(\bm{D}_{g’,1,t-1}) | \bm{D}].$$
\end{hyp}
Assumption \ref{hyp:strong_exogeneity2} is stronger than Assumption \ref{hyp:strong_exogeneity}. To simplify the remainder of the discussion, let us assume that treatment is binary. Then, combined with the standard parallel-trends assumption on groups' never-treated outcome, Assumption \ref{hyp:strong_exogeneity2} implies that for all groups such that $D_{g,1}=1$,
\begin{align}
&E\left[Y_{g,t}(\bm{1}_t) - Y_{g,t-1}(\bm{1}_{t-1})|\bm{D}\right]=E\left[Y_{g,t}(\bm{0}_t) - Y_{g,t-1}(\bm{0}_{t-1})|\bm{D}\right],\nonumber\\
\Longleftrightarrow \; &  E\left[Y_{g,t}(\bm{1}_t) - Y_{g,t}(\bm{0}_t)|\bm{D}\right]=E\left[Y_{g,t-1}(\bm{1}_{t-1})- Y_{g,t-1}(\bm{0}_{t-1})|\bm{D}\right].\label{eq:constanteffectsovertime}
\end{align}
\eqref{eq:constanteffectsovertime} means that in initially treated groups, the effect of being treated for $t$ periods should be the same as the effect of being treated for $t-1$ periods. This is an unpalatable restriction: it fails whenever lagged treatments affect the outcome and/or treatment effects are time-varying.
By contrast, when combined with the standard parallel-trends assumption on groups' never-treated outcome, Assumption \ref{hyp:strong_exogeneity} implies that for all groups such that $D_{g,1}=1$,
$$E(Y_{g,t}(\bm{1}_t) - Y_{g,t}(\bm{0}_t)|\bm{D})-(E(Y_{g,t-1}(\bm{1}_{t-1})- Y_{g,t-1}(\bm{0}_{t-1})|\bm{D})) \text{ does not vary across }g.$$
This means that the incremental effect of one treatment period should be the same in every initially treated group. This is of course a strong restriction, but unlike \eqref{eq:constanteffectsovertime}, it does not rule out effects of lagged treatments on the outcome, and it allows for time-varying effects.\footnote{\cite{deChaisemartin15b} and \cite{dCdH_AER} had already noted that in a model assuming away effects of lagged treatments on the outcome, parallel trends across groups with different baseline treatments essentially rules out time-varying effects.}

\paragraph{Comparison with some conditional event-study estimators of \cite{callaway2018}.}
When groups' period-one treatment varies and groups' treatment is always weakly larger than their baseline treatment, $\DID_\ell$ is equivalent to the estimator obtained by computing the \cite{callaway2018} estimator with the binarized and staggerized treatment, controlling non-parametrically for groups' baseline treatments. Despite this numerical equivalence, \cite{callaway2018}  consider binary and staggered treatments and do not suggest controlling for groups' period-one treatment: highlighting the importance of controlling for groups' baseline treatment is a contribution of this paper. Finally, when groups' period-one treatment varies and groups can experience a strictly lower treatment than their baseline treatment, $\DID_\ell$ is not equivalent to a version of the \cite{callaway2018} estimator.

\subsection{Normalized actual-versus-status-quo and event-study effects}\label{sub:AvsSQnorm}

\paragraph{Definition and estimation of the normalized actual-versus-status-quo effects.} For any $g$ such that $F_g\leq T_g$ and any $\ell \in \{1,...,T_g-F_g+1\}$, let
$$\delta^D_{g,\ell}=\sum_{k=0}^{\ell-1} (D_{g,F_g+k}-D_{g,1})$$
be the difference between the total treatment dose received by group $g$ from $F_g$ to $F_g-1+\ell$, and the total treatment dose it would have received in the status-quo counterfactual. Then, let
$$\delta^n_{g,\ell}=\frac{\delta_{g,\ell}}{\delta^D_{g,\ell}}.$$
We refer to $\delta^n_{g,\ell}$ as the normalized AVSQ effect of group $g$ at $F_g-1+\ell$. It follows directly from Lemma \ref{lem:id_delta_g_ell} that $\DID_{g,\ell}/\delta^D_{g,\ell}$ is conditionally unbiased for $\delta^n_{g,\ell}$.

\paragraph{$\delta^n_{g,\ell}$ is a weighted average of the effects the current and $\ell-1$ first treatment lags.}
For $k\in \{0,...,\ell-1\}$, let\footnote{We use the convention that for $k=0$ and any $d$, $\bm{d}_k$ stands for the empty vector. Accordingly when
 $k=\ell-1$, $(D_{g,F_g},..., D_{g,F_g-1+\ell-k-1})$ also stands for the empty vector. We sometimes refer to a cell's current treatment as its 0th treatment lag, and use the convention 0/0=0.}
\begin{align*}
s_{g,\ell,k}=&E\left[Y_{g,F_g-1+\ell}(\bm{D}_{g,1,F_g-1},D_{g,F_g},...,D_{g,F_g-1+\ell-k-1},\underline{D_{g,F_g-1+\ell-k}},\bm{D}_{g,1,k})\right.\nonumber\\
&\left. -\; Y_{g,F_g-1+\ell}(\bm{D}_{g,1,F_g-1},D_{g,F_g},..., D_{g,F_g-1+\ell-k-1},\underline{D_{g,1}}, \bm{D}_{g,1,k})
|\bm{D}\right]/(D_{g,F_g-1+\ell-k}-D_{g,1})
\end{align*}
be the slope of the expected potential outcome function of group $g$ at $F_g-1+\ell$ with respect to its $k$th treatment lag (the underlined term), when that lag is switched from its status-quo counterfactual value $D_{g,1}$ to its actual value $D_{g,F_g-1+\ell-k}$, whereas all its previous treatments are held at their actual values, and all its subsequent treatments are held at their status-quo value. For any $k\in \{0,...,\ell-1\}$, let
$$w_{g,\ell,k}=\frac{D_{g,F_g-1+\ell-k}-D_{g,1}}{\delta^D_{g,\ell}}.$$

\begin{lem}\label{lem:norm_delta_g_ell}
For every $(g,\ell)$ such that $1 \le \ell \le T_g-F_g+1$, $\delta^n_{g,\ell}=\sum_{k=0}^{\ell-1} w_{g,\ell,k} s_{g,\ell,k}$. Moreover, $\sum_{k=0}^{\ell-1} w_{g,\ell,k}=1$ for all $(g,\ell)$ and under Restriction \ref{hyp:monotonic_design}, $w_{g,\ell,k}\geq 0$.
\end{lem}
Lemma \ref{lem:norm_delta_g_ell} shows that $\delta^n_{g,\ell}$ is a weighted average of the slopes of $g$'s potential outcome at $F_g-1+\ell$ with respect to its $\ell-1$ first treatment lags, where for $k \in \{0,...,\ell-1\}$, the effect of the $k$th lag receives a weight proportional to the absolute value of the difference between $g$'s $k$th treatment lag and its status-quo treatment.

\paragraph{Lemma \ref{lem:norm_delta_g_ell} in some specific designs.}
For concreteness, we rewrite the result in Lemma \ref{lem:norm_delta_g_ell} in three specific designs.
First, in binary and staggered designs,
Lemma \ref{lem:norm_delta_g_ell} reduces to
$$\delta^n_{g,\ell}=\frac{1}{\ell}\sum_{k=0}^{\ell-1} E\left[Y_{g,F_g-1+\ell}(\0_{F_g-1},\1_{\ell-k-1},1,\0_{k})-Y_{g,F_g-1+\ell}(\0_{F_g-1},\1_{\ell-k-1},0,\0_{k})
|\bm{D}\right].$$
Then, $\delta^n_{g,\ell}$ is the simple average, across $k$ ranging from $0$ to $\ell-1$, of the effect of switching the $k$th treatment lag from $0$ to $1$, holding previous treatments at $1$ and subsequent treatments at $0$.
Second, in the one-shot-treatment designs defined in \eqref{eq:oneshottreatment}, Lemma \ref{lem:norm_delta_g_ell} reduces to
$$\delta^n_{g,\ell}=E\left[Y_{g,F_g-1+\ell}(\0_{F_g-1},1,\0_{\ell-1})-Y_{g,F_g-1+\ell}(\0_{F_g-1},0,\0_{\ell-1})
|\bm{D}\right],$$
so $\delta^n_{g,\ell}$ is just the effect of switching the $\ell-1$th treatment lag from $0$ to $1$, holding all other treatments at $0$.
Third, in Design \ref{des:stag_nonbinary} ($D_{g,t}=I_g1\{t\geq F_g\}$), Lemma \ref{lem:norm_delta_g_ell} reduces to
$$\delta^n_{g,\ell}=\frac{1}{\ell}\sum_{k=0}^{\ell-1} \frac{E\left[Y_{g,F_g-1+\ell}(\0_{F_g-1},\bm{I}_{g,\ell-k-1},I_g,\0_{k})-Y_{g,F_g-1+\ell}(\0_{F_g-1},\bm{I}_{g,\ell-k-1},0,\0_{k})
|\bm{D}\right]}{I_g}.$$
Thus, in staggered designs with group-specific treatment intensities, $\delta^n_{g,\ell}$ is the average, across $k$ ranging from $0$ to $\ell-1$, of the effect of switching the $k$th lag from $0$ to $I_g$, normalized by $I_g$.

\paragraph{Definition and estimation of the normalized event-study effects.}
Let $\delta^D_{\ell}=\frac{1}{N_\ell}\sum_{g:F_{g}-1+\ell\leq T_{g}}|\delta^D_{g,\ell}|$.
For $\ell \in \{1,...,L\}$, let
$$\delta^n_{\ell}=\frac{1}{N_\ell}\sum_{g:F_g-1+\ell\leq T_g}\frac{|\delta^D_{g,\ell}|}{\delta^D_{\ell}}\delta^n_{g,\ell}.$$
Note the following relation between the non-normalized and normalized event-study effects:
\begin{equation}
\delta^n_{\ell}=\frac{\delta_{\ell}}{\delta^D_{\ell}}.	
	\label{eq:relationship_norm_unnormeffects}
\end{equation}
$\delta^n_{\ell}$ is a weighted average of all the $\delta^n_{g,\ell}$ parameters that can be estimated, with weights proportional to $|\delta^D_{g,\ell}|$. In designs where some $g$s are such that $|\delta^D_{g,\ell}|$ is close to zero, the estimator of the unweighted average of the $\delta^n_{g,\ell}$s may be very noisy. This is what leads us to consider the weighted average above.
It follows directly
from Lemma \ref{lem:id_delta_g_ell} that
$$\DID^n_{\ell}:=\frac{1}{N_\ell}\sum_{g:F_g-1+\ell\leq T_g}\frac{|\delta^D_{g,\ell}|}{\delta^D_{\ell}}\frac{\DID_{g,\ell}}{\delta^D_{g,\ell}}$$
is conditionally unbiased for $\delta^n_{\ell}$. Similarly to \eqref{eq:relationship_norm_unnormeffects},
$\DID^n_{\ell}=\DID_{\ell}/\delta^D_{\ell}.$

\paragraph{Interpretation of $\delta^n_{\ell}$.} It follows from Lemma \ref{lem:norm_delta_g_ell} that $\delta^n_{\ell}$ is a weighted average of the effects of groups' current and $\ell-1$ first treatment lags on their outcome. The total weight assigned by $\delta^n_{\ell}$ to the effect of the $k$th-lag (for $0\le k\le \ell-1$) is equal to
\begin{align*}
w_{\ell,k}=&\frac{1}{N_\ell}\sum_{g:F_g-1+\ell\leq T_g}\frac{|D_{g,F_g-1+\ell-k}-D_{g,1}|}{\delta^D_{\ell}}.
\end{align*}
In the one-shot treatment designs defined in \eqref{eq:oneshottreatment}, $w_{\ell,\ell-1}=1$, so $\delta^n_{\ell}$ only measures the effect of the $\ell-1$th treatment lag. In more complicated designs, $\delta^n_{\ell}$ averages together the effect of different treatment lags. When groups' treatment can only change once, $w_{\ell,k}=1/\ell$, so $\ell \mapsto w_{\ell,k}$ is decreasing for $\ell\geq k+1$: $\delta^n_{\ell}$ assigns less weight to the effect of recent treatments when $\ell$ increases. Finally, one always has $w_{1,0}=1$ and  $w_{\ell,0}\leq 1$ for $\ell \ge 2$: $\delta^n_{1}$ only averages effects of groups' current treatment, whereas $\delta^n_{\ell}$ also averages effects of treatment lags for $\ell \ge 2$. We recommend reporting $k \mapsto w_{\ell,k}$, to document which lags contribute the most to $\delta^n_{\ell}$.\footnote{One may use $\ell\mapsto\delta^n_{g,\ell}$ to separately estimate the effect of each treatment lag, if one assumes that lags' effects are additively separable, linear, and constant over time, as we had shown in a previous version of this paper \citep[see Section 4 of][]{chaisemartin2020intertemporal}.}

\subsection{Cost-benefit analysis}\label{sub:cost-benefit}

\paragraph{Assumption on the design.} In this section, we strengthen Restriction \ref{hyp:monotonic_design}.
\begin{conddes}\label{hyp:increasing_design}
(Lowest treatment at period one) $\forall g\in \{1,...,G\}$, $D_{g,t}\geq D_{g,1}$ $\forall t$.
\end{conddes}
We impose Restriction \ref{hyp:increasing_design} to reduce the notational burden. When it fails, one can just conduct the cost-benefit analysis separately for groups with $S_g=1$ and for groups with $S_g=-1$, as was done in a previous version of this paper \citep[see][]{chaisemartin2020intertemporal}.

\paragraph{Cost-benefit analysis.}
In this section, our target parameter is
$$\delta:=\frac{\sum_{g:F_g\leq T_g}\sum_{\ell=1}^{T_g-F_g+1}\delta_{g,\ell}}{\sum_{g:F_g\leq T_g}\sum_{\ell=1}^{T_g-F_g+1}(D_{g,F_g-1+\ell}-D_{g,1})}.$$
To motivate it, we take the perspective of a planner, seeking to conduct a cost-benefit analysis comparing groups' actual treatments $\bm{D}$ to the counterfactual ``status-quo'' scenario where they would have kept throughout their period-1 treatment. We assume that the outcome can be converted into a monetary equivalent, or at least that treatment effects per USD spent can be used to determine if the treatment is worthwhile. Then, $\delta_{g,\ell}$ is the expected benefit or loss, in group $g$ and at period $F_g-1+\ell$, of having received the actual rather than the status-quo treatments since period $F_g$, the first period when group $g$ deviated from its status-quo treatment. We also assume that the treatment's cost is linear in the dose administered, and we let $c_{g,\ell}\geq 0$ denote the cost of administering one treatment unit in group $g$ at period $F_g-1+\ell$. To simplify, we assume that $c_{g,\ell}$ is non-stochastic, and that it can be readily computed, for instance using the accounts of the organization delivering the treatment. Assuming that the planner's discount factor is equal to $1$, groups' actual treatments are beneficial in monetary terms relative to the status quo, up to period $T_g$, if and only if
$$\sum_{g:F_g\leq T_g}\sum_{\ell=1}^{T_g-F_g+1}\delta_{g,\ell}-\sum_{g:F_g\leq T_g}\sum_{\ell=1}^{T_g-F_g+1}c_{g,\ell}(D_{g,F_g-1+\ell}-D_{g,1})> 0
\Leftrightarrow \delta > c,$$
where $$c=\frac{\sum_{g:F_g\leq T_g}\sum_{\ell=1}^{T_g-F_g+1}c_{g,\ell}(D_{g,F_g-1+\ell}-D_{g,1})}{\sum_{g:F_g\leq T_g}\sum_{\ell=1}^{T_g-F_g+1}(D_{g,F_g-1+\ell}-D_{g,1})}$$
is the average treatment cost, across all the incremental treatment doses
received with respect to the status-quo counterfactual.

\paragraph{Estimating $\delta$.} It follows directly  from Lemma \ref{lem:id_delta_g_ell} that
$$\widehat{\delta}:=\frac{\sum_{g:F_g\leq T_g}\sum_{\ell=1}^{T_g-F_g+1}\DID_{g,\ell}}{\sum_{g:F_g\leq T_g}\sum_{\ell=1}^{T_g-F_g+1}(D_{g,F_g-1+\ell}-D_{g,1})}$$
is conditionally unbiased for $\delta$.

\paragraph{Formula for $\delta$ in binary and staggered designs.}
In binary and staggered designs, one has
$$\delta=E\left[\frac{1}{\sum_{(g,t)}D_{g,t}}\sum_{(g,t):D_{g,t}=1}\left(Y_{g,t}\left(\0_{F_g-1},\1_{t-F_g+1}\right)-Y_{g,t}\left(\0_t\right)\right)\middle| \bm{D}\right],$$
if there is at least one never-treated group. Then, $\delta$ is
the average effect of having been treated for $t-F_g+1$ periods across all treated $(g,t)$ cells, a parameter that generalizes the average treatment effect on the treated to our setting with dynamic effects.

\paragraph{Interpretation of $\delta$ as an average total effect per unit of treatment.} Let us first consider a simple example with two groups and four periods, such that $D_{1,1}=0$, $D_{1,2}=1$, $D_{1,3}=1$, and $D_{1,4}=0$, while group $2$ is never treated. Then,
\begin{align}
\delta&=\frac{E\left[Y_{1,2}(0,1)-Y_{1,2}(0,0)+Y_{1,3}(0,1,1)-Y_{1,3}(0,0,0)+Y_{1,4}(0,1,1,0)-Y_{1,4}(0,0,0,0)|\bm{D}\right]}{1+1+0}\notag \\
&=\frac{1}{2} E\left[Y_{1,2}(0,1)-Y_{1,2}(0,0)+Y_{1,3}(0,1,0)-Y_{1,3}(0,0,0)+Y_{1,4}(0,1,0,0)-Y_{1,4}(0,0,0,0)|\bm{D}\right] \notag \\
&+\frac{1}{2}E\left[Y_{1,3}(0,1,1)-Y_{1,3}(0,1,0)+Y_{1,4}(0,1,1,0)-Y_{1,4}(0,1,0,0)|\bm{D}\right].\label{eq:delta_plus_total_eff}
\end{align}
The first expectation in \eqref{eq:delta_plus_total_eff} is the total effect produced by group $1$'s period-$2$ treatment, at periods $2$, $3$, and $4$, relative to the situation where it would have always remained untreated. The second expectation in \eqref{eq:delta_plus_total_eff}  is the total effect produced by group $1$'s period-$3$ treatment, at periods $3$ and $4$, conditional on its period-$2$ treatment and relative to the situation where it would have been untreated at periods $3$ and $4$. Accordingly, $\delta$ is the average \textit{total} effect of those two treatments. A similar interpretation holds beyond this simple example. For $k\in \{0,...,\ell-1\}$, let $\delta_{g,\ell,k}$ be the numerator of the slope $s_{g,\ell,k}$ defined above.
$\delta_{g,\ell,k}$ is the effect, on the expected potential outcome of group $g$ at $F_g-1+\ell$, of switching its $k$th treatment lag from its status-quo to its actual value, whereas all its previous treatments are held at their actual values, and all its subsequent treatments are held at their status-quo value.
As $\delta_{g,\ell}=\sum_{k=0}^{\ell-1} \delta_{g,\ell,k}$,
\begin{align*}
\delta=&\frac{\sum_{g:F_g\leq T_g}\sum_{\ell=1}^{T_g-F_g+1}\sum_{k=0}^{\ell-1} \delta_{g,\ell,k}}{\sum_{g:F_g\leq T_g}\sum_{\ell=1}^{T_g-F_g+1}(D_{g,F_g-1+\ell}-D_{g,1})}=\frac{\sum_{g:F_g\leq T_g}\sum_{k=0}^{T_g-F_g}\sum_{\ell=k+1}^{T_g-F_g+1} \delta_{g,\ell,k}}{\sum_{g:F_g\leq T_g}\sum_{k=0}^{T_g-F_g}(D_{g,F_g+k}-D_{g,1})}.
\end{align*}
$\sum_{\ell=k+1}^{T_g-F_g+1} \delta_{g,\ell,k}$ is the total effect, from period $F_g+k$ to $T_g$, of switching $g$'s period $F_g+k$ treatment from $D_{g,1}$ to $D_{g,F_g+k}$. The sum of total effects in the numerator of $\delta$ is scaled by the sum of all the incremental treatments $D_{g,F_g+k}-D_{g,1}$ that generate those total effects. Accordingly, $\delta$ may be interpreted as an average total effect per unit of treatment.

\paragraph{Average number of time periods over which the effect of a dose is cumulated.}
The effect of switching $g$'s period $F_g+k$ treatment from $D_{g,1}$ to $D_{g,F_g+k}$ is cumulated from period $F_g+k$ to $T_g$, namely over $T_g-F_g-k+1$ periods. To interpret $\delta$, one may compute
\begin{align*}
\frac{\sum_{g:F_g\leq T_g}\sum_{k=0}^{T_g-F_g}(D_{g,F_g+k}-D_{g,1})(T_g-F_g-k+1)}{\sum_{g:F_g\leq T_g}\sum_{k=0}^{T_g-F_g}(D_{g,F_g+k}-D_{g,1})},
\end{align*}
the average number of time periods over which the effect of a dose is cumulated, on average across all incremental doses received by switchers over the study period. One may then divide $\delta$ by this average number of time periods, to get an average effect of being exposed to one dose of treatment for one period.

\paragraph{Connection between $\delta$ and the event-study effects $\delta_{\ell}$.}
For $\ell \in \{1,...,L\}$, let
$w_{\ell}=N_{\ell}/\sum_{g:F_g\leq T_g}\sum_{\ell'=1}^{T_g-F_g+1}(D_{g,F_g-1+\ell'}-D_{g,1})$.
\begin{lem}\label{lem_main}
If Restrictions \ref{hyp:non_pathological_design} and \ref{hyp:increasing_design} hold,
$\delta = \sum_{\ell=1}^{L} w_{\ell} \delta_{\ell}$.
\end{lem}
Lemma \ref{lem_main} shows that $\delta$ is equal to a linear combination, with non-negative factors, of the non-normalized event-study effects $(\delta_{\ell})_{1\leq \ell \leq L}$.

\subsection{Inference}  
\label{sec:inference}

\paragraph{Assumptions.} Our asymptotic results are based on the three assumptions below. Hereafter, let $\Sigma_g=V(\bm{Y}_g|\bm{D})$. For any symmetric matrix $\Sigma$, $\underline{\rho}(\Sigma)$ denotes its smallest eigenvalue. Let $$\mathcal{L}=\{\ell:\lim_{G\to\infty} N_{\ell}=\infty \text{ almost surely}\}$$ be the set of values of $\ell$ such that the number of groups for which $\delta_{g,\ell}$ can be estimated goes to infinity almost surely (a.s.) as $G$ tends to infinity.
\begin{hyp}
\label{hyp:indep_groups}
(Independent groups) Conditional on $(\bm{D}_g)_{g\ge 1}$, the vectors $(\bm{Y}_g)_{g\geq 1}$ are mutually independent.
\end{hyp}

\begin{hyp}
\label{hyp:asym_design}
	(Asymptotic restriction on the design)
\begin{enumerate}
\item $\limsup_G \# \mathcal{D}^{\text{r}}_1<\infty$ almost surely (a.s.).
\item $\mathcal{L}\ne \emptyset$.
\item  For $\ell \in \mathcal{L}$, let $v^G_{d,s,\ell} := \#\{g\le G: D_{g,1}=d, S_g=s, F_g-1+\ell \le T_g \}$. For all $(d, s)$ such that $\lim_G v^G_{d,s,\ell}=\infty$ a.s., we have
$$\liminf_G \frac{\#\{g\le G: D_{g,1}=d, F_g = \max_{g’: D_{g’,1}=d} F_{g’}\}}{v^G_{d,s,\ell}} > 0 \quad \text{a.s.}$$
\end{enumerate}
  \end{hyp}

\begin{hyp}
\label{hyp:moment_cond}
(Regularity conditions) We have, for some $\delta>0$ and a.s.,
$$\sup_{g\ge 1, t\in\{1,...,T\}} E[|Y_{g,t}|^{2+\delta}|\bm{D}]<\infty \;\; \text{and }\; \inf_{g\ge 1} \underline{\rho}\left(\Sigma_g\right) > 0.$$
\end{hyp}
Assumption \ref{hyp:indep_groups} allows for serial correlation of the treatments and outcomes within each group, which are important features to account for in DID studies \citep{bertrand2004}. Assumption \ref{hyp:indep_groups} is also weaker than the common assumption that $(\bm{D}_g,\bm{Y}_g)_{g\ge 1}$ are i.i.d. or even simply independent.  In particular, it allows for any form of dependence between groups' treatments $(\bm{D}_g)_{g\ge 1}$. Assumption \ref{hyp:asym_design} may be seen as a reinforcement and an asymptotic version of Restriction \ref{hyp:non_pathological_design}. It is stated as an assumption because it is not directly verifiable from the data. Point 1 requires that the number of values of the baseline treatment remains finite when the number of groups goes to infinity. Point 2 requires that there is at least one value of $\ell$ for which the number of groups such that $\delta_{g,\ell}$ can be estimated goes to $\infty$ a.s., which is necessary to have that $\DID_\ell$ is consistent. With i.i.d. groups, $N_\ell/G$ converges almost surely to a positive constant for all $\ell \in \mathcal{L}$. Instead, we do not impose any restriction on the rate at which $N_\ell$ goes to infinity. Point 3 requires that if the number of groups $g$ such that (i) $\delta_{g,\ell}$ can be estimated and (ii) $(D_{g,1},S_g)=(d,s)$ goes to infinity, then the number of groups $g'$ with $D_{g',1}=d$ that switch the latest (or that never switch) also tends to infinity at the same rate. Otherwise, some control groups intervening in $\DID_\ell$ may have a weight tending to infinity, which could make asymptotic normality of $\DID_\ell$ fail.  Noteworthy, Point 3 of Assumption \ref{hyp:asym_design} automatically holds if groups' treatment paths $(\bm{D}_g)_{g\ge 1}$ are i.i.d. Finally, Assumption \ref{hyp:moment_cond} ensures in particular that we can apply the Lyapunov central limit theorem in our set-up with independent but not identically distributed variables. The second condition therein prevents degenerate situations where some non-trivial linear combinations of the $(Y_{g,1},...,Y_{g,T})$ in $\DID_\ell$ would actually be constant.

\paragraph{Asymptotic results.} We now establish the asymptotic properties of $\DID_{\ell}$  when the number of groups tends to infinity. We also propose a confidence interval for $\delta_\ell$ and show its asymptotic validity. To define this confidence interval, let $N^g_{t,\ell}=\sum_{g'\le G:D_{g',1}=D_{g,1}} S_{g'} \ind{F_{g'}=t-\ell+1}$ and
\begin{align*}
\lambda^G_{g,\ell,t} = & S_g\ind{F_g\le T_g-\ell+1}\left(\ind{F_g=t-\ell+1} - \ind{F_g=t+1}\right) -\frac{N^g_{t,\ell}}{N^g_t}\ind{F_g>t}  \\
+ & \frac{N^g_{t+\ell,\ell}}{N^g_{t+\ell}}\ind{F_g>t+\ell}.
\end{align*}
where, again, we use the convention that $0/0=0$. Then, define $\bm{\lambda}^G_{g,\ell} =(\lambda^G_{g,\ell,1},\cdots,\lambda^G_{g,\ell,T})$, $\bm{Y}_g=(Y_{g,1},\cdots,Y_{g,T})'$ and $U^G_{g,\ell} = \bm{\lambda}^G_{g, \ell}\bm{Y}_g$. Some algebra shows that
$$\DID_\ell = \frac{1}{N_\ell}\sum_{g=1}^G U^G_{g,\ell}.$$
As a result, under Assumption \ref{hyp:indep_groups},
$$V(N_\ell^{1/2} \DID_\ell|\bm{D}) = \frac{1}{N_\ell}\sum_{g=1}^G E\left[(U^G_{g,\ell} - E(U^G_{g,\ell}|\bm{D}))^2|\bm{D}\right].$$
We then consider an estimator of $V(N_\ell^{1/2} \DID_\ell|\bm{D})$ of the form
$$\frac{1}{N_\ell}\sum_{g=1}^G \left(U^G_{g,\ell} - \widehat{\theta}_g\right)^2,$$
where $\widehat{\theta}_g$ is an estimator of $E(U^G_{g,\ell}|\bm{D})$. A difficulty is that in our i.n.i.d set-up, we cannot estimate consistently $E(U^G_{g,\ell}|\bm{D})$, as this expectation varies across groups. As a result, inference is in general conservative, whichever estimators $(\widehat{\theta}_g)_{g\ge 1}$ we consider. However, some estimators $(\widehat{\theta}_g)_{g\ge 1}$ can lead to non-conservative inference under some assumptions. 
To define these $(\widehat{\theta}_g)_{g\ge 1}$, let us remark that under Assumption \ref{hyp:asym_design}, the number of distinct values of $(D_{g,1},F_g,S_g)_{g: D_{g,1}\in \mathcal{D}^{\text{r}}_1}$ is finite;\footnote{Groups $g$ such that $D_{g,1}\not\in \mathcal{D}^{\text{r}}_1$ do not affect $\DID_\ell$ (since $\bm{\lambda}^G_{g, \ell}= \bm{0}_T$) and thus can be ignored.} we can thus write this set as $\{(d_k, f_k, s_k):\ k=1,...,K\}$. Then, let $\mathcal{C}_k=\{g\ge 1:D_{g,1}=d_k, F_g=f_k, S_g=s_k\}$ and $\mathcal{C}^G_k=\mathcal{C}_k\cap\{1,...,G\}$. The $(\mathcal{C}^G_k)_{k=1,...,K}$ are called cohorts hereafter. If $g\in \mathcal{C}^G_k$, we define $\widehat{E}(U^G_{g,\ell}|\bm{D})= (1/\#\mathcal{C}^G_k) \sum_{g'\in \mathcal{C}^G_k} U^G_{g',\ell}$, and we let
$$\widehat{\sigma}^2_{\ell}  = \frac{1}{N_\ell}\sum_{g=1}^G \left(U^G_{g,\ell} - \widehat{E}(U^G_{g,\ell}|\bm{D})\right)^2.$$
The confidence interval of nominal level $1-\alpha$ on $\delta_\ell$ that we consider is then
$$\CI_{1-\alpha} = \left[\DID_\ell \pm z_{1-\alpha/2} \frac{\widehat{\sigma}_{\ell}}{N_\ell^{1/2}} \right],$$
where $z_{1-\alpha/2}$ is the quantile of order $1-\alpha/2$ of the standard normal distribution.

\begin{thm}\label{thm:asym}
	Suppose that Assumptions \ref{hyp:no_antic}-\ref{hyp:strong_exogeneity} and \ref{hyp:indep_groups}-\ref{hyp:moment_cond} hold. Then, for all $\ell\in\mathcal{L}$, conditional on $(\bm{D}_g)_{g\geq 1}$ and almost surely,
	\begin{align*}
		\DID_{\ell} - \delta_{\ell}& \convP 0, \\
		\sqrt{N_\ell}\frac{\DID_{\ell} - \delta_{\ell}}{\left(\frac{1}{N_\ell}\sum_{g=1}^G V(U^G_{g,\ell}|\bm{D})\right)^{1/2}}  & \convL \mathcal{N}(0,1), \\
		\liminf_{G\to\infty} \Pr\left[\delta_\ell\in\CI_{1-\alpha}|\bm{D}\right] & \ge 1-\alpha,
	\end{align*}
	with an equality if we assume, instead of Assumption \ref{hyp:indep_groups}, that $(\bm{D}_g, \bm{Y}_g)_{g\ge 1}$ are i.i.d. and $\bm{D}_g$ is a function of $(D_{g,1}, F_g, S_g)$.
\end{thm}

Theorem \ref{thm:asym} shows that for almost all realizations of $(\bm{D}_g)_{g\geq 1}$, $\DID_\ell$ is asymptotically normal, and the confidence interval of $\delta_\ell$ is asymptotically conservative. Moreover, if groups are identically distributed and $\bm{D}_g$ is a function of $(D_{g,1}, F_g, S_g)$, the confidence interval reaches its nominal level asymptotically. This latter condition holds if the treatment is binary and can change only once, as is for instance the case in Design \ref{des:binaryandstaggered}. Finally, the results of Theorem \ref{thm:asym} easily extend to the normalized parameter $\delta^n_\ell$: recall that
  $\delta^n_\ell=\delta_\ell/\delta^D_\ell$ and $\DID^n_\ell=\DID_\ell/\delta^D_\ell$, where  $\delta^D_\ell$ is a function of $\bm{D}$ only. Similar asymptotic results could also be obtained for the cost-benefit parameter $\delta$.

\subsection{Extensions} 
\label{sub:extensions}

In this section we briefly review some extensions, fully covered in our Web Appendix (WA).

\paragraph{Placebo estimators (Section \ref{sub:placebo_estimators} of the WA).}\label{sec:placebos}

We propose placebo estimators one can use to test Assumptions \ref{hyp:no_antic} and \ref{hyp:strong_exogeneity}. For any $g:3\leq F_g\leq T_g$ and $\ell \in \{1,...,\min(T_g-F_g+1,F_g-2)\}$,
\begin{equation*}
\DID^{\pl}_{g,\ell}:=Y_{g,F_g-1-\ell}-Y_{g,F_g-1} -
\frac{1}{N^g_{F_g-1+\ell}}\sum_{g':D_{g',1}=D_{g,1},F_{g'} >F_g-1+\ell}(Y_{g',F_g-1-\ell}-Y_{g',F_g-1}).
\end{equation*}
$\DID^{\pl}_{g,\ell}$ mimicks $\DID_{g,\ell}$. Like $\DID_{g,\ell}$, it compares the outcome evolution of $g$ to that of groups with the same baseline treatment as $g$, and that have kept that treatment from period $1$ to $F_g-1+\ell$. But unlike $\DID_{g,\ell}$, it compares those groups' outcome evolutions from period $F_g-1-\ell$ to period $F_g-1$, namely before group $g$'s treatment changes for the first time. Accordingly, $\DID^{\pl}_{g,\ell}$ assesses if $g$ and groups whose treatment has not changed yet at $F_g-1+\ell$ experience the same evolution of their status-quo outcome over $\ell$ periods, the number of periods over which parallel trends has to hold for $\DID_{g,\ell}$ to be unbiased for $\delta_{g,\ell}$. One can show that under Assumptions \ref{hyp:no_antic} and \ref{hyp:strong_exogeneity}, $E(\DID^{\pl}_{g,\ell}|\bm{D})=0$. Whereas $\DID_{g,\ell}$ goes from the past (period $F_g-1$) to the future (period $F_g-1+\ell$), $\DID^{\pl}_{g,\ell}$ goes from the future (period $F_g-1$) to the past (period $F_g-1-\ell$). This follows the standard practice in event-study regressions, where the reference period is the one before the event. In Section \ref{sub:placebo_estimators} of the WA, we define placebo estimators $\DID^{\pl}_{\ell}$ mimicking $\DID_{\ell}$. Essentially, the placebos replace $\DID_{g,\ell}$ by $\DID^{\pl}_{g,\ell}$ in the estimators' definitions.

\paragraph{Controlling for covariates (Section \ref{sub:covariates} of the WA).}

With covariates, we replace the equality in Assumption \ref{hyp:strong_exogeneity} by
\begin{align}
& E[Y_{g,t}(\bm{D}_{g,1,t}) - Y_{g,t-1}(\bm{D}_{g,1,t-1}) - (X_{g,t} - X_{g,t-1})'\theta_{D_{g,1}} | \bm{D},\bm{X}] \notag\\
= & E[Y_{g’,t}(\bm{D}_{g’,1,t}) - Y_{g’,t-1}(\bm{D}_{g’,1,t-1}) - (X_{g',t} - X_{g',t-1})'\theta_{D_{g',1}} | \bm{D},\bm{X}], \label{eq:PT_X}
\end{align}
where $\bm{X}$ stacks the covariates of all groups at all periods. \eqref{eq:PT_X} means that groups can experience differential trends, provided those differential trends are fully explained by changes in their covariates. Then, we estimate $\theta_d$ by regressing $Y_{g,t}-Y_{g,t-1}$ on $X_{g,t}-X_{g,t-1}$ in the sample of $(g,t)$s such that $D_{g,1}=d$ and $F_g>t$. Finally, we define similar estimators of $\delta_{\ell}$, $\delta^n_{\ell}$ and $\delta$ as above, replacing $Y_{g,t}-Y_{g,t-1}$ by $Y_{g,t}-Y_{g,t-1}-(X_{g,t}-X_{g,t-1})'\widehat{\theta}_{D_{g,1}}$.
This approach can also be used to control for time-invariant covariates, defining $X_{g,t}=X_g\times t$. Then, $(X_{g,t} - X_{g,t-1})'\theta_{D_{g,1}}=X_{g}'\theta_{D_{g,1}}$, so \eqref{eq:PT_X} reduces to a conditional parallel-trends assumption, with a linear time-invariant functional form for the conditional counterfactual trend. Alternatively, one may define $X_{g,t}$ as the interaction of $X_g$ and period fixed effects, in which case \eqref{eq:PT_X} reduces to a conditional parallel-trends assumption, with a period-specific linear functional form for the conditional trend.

\paragraph{Other extensions (Sections \ref{sub:grp_spec_lin} to \ref{sub:continuous} of the WA)} 

First, we show how to allow for group-specific linear trends. Second, we show how to allow for different trends across sets of groups defined by their value of a time-invariant covariate $X_g$. For instance, one may want to allow for state-specific trends in a county-level analysis.
Third, we show how to estimate heterogeneous treatment effects.
Fourth, in Design \ref{des:binaryoneexit}, we propose an extension of our estimators, to separately estimate the effects of joining and leaving treatment.
Fifth, we discuss how our estimators can be used in fuzzy designs, where the treatment varies within $(g,t)$ cells \citep[][]{deChaisemartin15b}.
Sixth, ruling out the effect of past treatments beyond $k$ lags allows us to propose a solution to the initial-conditions problem. Seventh, we show how the normalized event-study effects can be used to test the null that the current and lagged treatments all have the same effects. Finally, we build upon ideas initially proposed by \cite{de2022difference} to sketch an extension of our estimators to designs where
(i) in Restriction 1 fails, because groups' period-one treatment is continuous.

\section{Current Practice}
\label{sec:current_practice}

We conducted a census of highly-influential papers published by the AER from 2015 to 2019 and that have used TWFE regressions. To do so, we first ran a Google Scholar (GS) search of all papers published by the AER in 2015, sorted according to GS's relevance criteria, which is nearly equivalent to sorting on GS citations \citep{beel2009google}. We systematically reviewed the first 20 papers, and identified three that have estimated at least one TWFE regression. We repeated the same process for 2016, 2017, 2018, and 2019. In total, we reviewed 100 papers, and found 26 that have estimated at least one TWFE regression. The list of papers can be found in Table \ref{tab:survey} in the Web Appendix. Of those 26 papers, two have a binary treatment with no variation in treatment timing,\footnote{With a binary treatment and no variation in treatment timing, TWFE regressions estimate the average treatment effect on the treated, see \cite{de2022survey}.} and four have a binary treatment and a staggered adoption design. Four of those six papers estimate dynamic treatment effects, using the event-study regression considered by \cite{abraham2018}. Among the remaining 20 papers, 11 estimate only static TWFE regressions, thus implicitly ruling out dynamic treatment effects. The remaining nine papers estimate dynamic effects, using one of the three estimations methods described below.

\subsection{TWFE regressions with treatment intensity interacted with period FEs.}

\paragraph{Designs with variation in treatment intensity and no variation in timing.} Among the nine papers in our census that do not have a binary and staggered design and estimate dynamic effects, some are such that $D_{g,t}=I_g1\{t\geq F\}$, for $2\leq F\leq T$, a special case of Design \ref{des:stag_nonbinary} where groups all start getting treated at the same date $F$ with group-specific intensities $I_g$, where $I_g$ may be equal to zero if there are control groups that remain completely untreated. In such designs, researchers often estimate the following TWFE regression.
\begin{reg}\label{reg:twfe_d*t}
(TWFE regression with treatment intensity interacted with period FEs) For every $\ell \in \{-F+2,...,T-F+1\},\ell \ne 0$, let $\widehat{\beta}_{fe,\ell}$ be the coefficient on $I_g1\{t=F-1+\ell\},$ in a regression of $Y_{g,t}$ on group and period FEs and $\left(I_g1\{t=F-1+\ell\}\right)_{\ell \in \{-F+2,...,T-F+1\},\ell \ne 0}$.
\end{reg}
Intuitively, for $\ell \in \{1,...,T-F+1\}$, researchers hope that $\widehat{\beta}_{fe,\ell}$ estimates some average effect of increasing the treatment by one unit for $\ell$ periods. For $\ell \in \{-F+2,...,-1\}$, researchers hope that $\widehat{\beta}_{fe,\ell}$ can be used as a placebo estimator, to test Assumptions \ref{hyp:no_antic} and \ref{hyp:strong_exogeneity}.\footnote{Assumption \ref{hyp:strong_exogeneity} reduces to a parallel-trends assumption on the never-treated outcome in those designs.} Instead of Regression \ref{reg:twfe_d*t}, researchers have also estimated regressions of $Y_{g,F-1+\ell}-Y_{g,F-1}$ on $I_g$. One can show that the coefficient on $I_g$ in that regression is equal to $\widehat{\beta}_{fe,\ell}$.

\paragraph{A decomposition of the coefficients in TWFE regressions with treatment intensity interacted with period FEs.}
We now study what the coefficients $\widehat{\beta}_{fe,\ell}$ identify. First, note that when $D_{g,t}=I_g1\{t\geq F\}$, $\delta_{g,\ell}=E(Y_{g,F-1+\ell}(\0_{F-1},\bm{I}_{g,\ell})-Y_{g,F-1+\ell}(\0_{F-1+\ell})|\bm{D})$, the effect of having received $I_g$ rather than $0$ units of treatment for $\ell$ periods. Let $\bar{I}=(1/G)\sum_{g=1}^G I_g.$
\begin{prop}\label{prop:regression1}
If Assumptions \ref{hyp:no_antic}-\ref{hyp:strong_exogeneity} hold,  $D_{g,t}=I_g1\{t\geq F\}$ for $2\leq F\leq T$ and $I_g$ is not constant across groups, then:
\begin{enumerate}
\item $\forall$ $\ell \in \{1,...,T-F+1\}$,
$$E\left(\widehat{\beta}_{fe,\ell}|\bm{D}\right)=\sum_{g:I_g\ne 0}w^{fe}_{g}\frac{\delta_{g,\ell}}{I_g}, $$
where $$w^{fe}_{g}=\frac{I_g(I_g-\bar{I})}{\sum_{g':I_{g'}\ne 0}I_{g'}(I_{g'}-\bar{I})}.$$
\item If $F>2$, for all $\ell \in \{-F+2,...,-1\}$, $E\left(\widehat{\beta}_{fe,\ell}|\bm{D}\right)=0$.
\end{enumerate}
\end{prop}

\paragraph{Interpretation of, and some remarks on, Proposition \ref{prop:regression1}.} Point 1 of Proposition \ref{prop:regression1} shows that for $\ell \in \{1,...,T-F+1\}$, $\widehat{\beta}_{fe,\ell}$ estimates a weighted sum across groups of $\delta_{g,\ell}$, group $g$'s effect of having received $I_g$ rather than $0$ units of treatment for $\ell$ periods, normalized by $I_g$. The weights $w^{fe}_{g}$ sum to one but are not equal to $1/\#\{g:I_g\ne 0\}$, so $\widehat{\beta}_{fe,\ell}$ may be biased for the average of $\delta_{g,\ell}/I_g$ across the treated groups. Perhaps more worryingly, some of the weights may be negative: treatment effects of groups with a positive treatment intensity smaller than $\bar{I}$ are weighted negatively by $\widehat{\beta}_{fe,\ell}$. Then, $E\left(\widehat{\beta}_{fe,\ell}|\bm{D}\right)$ does not satisfy the no-sign-reversal property: it could be, say, negative, even if for all $(g,t)$, $(d_1,...,d_t)\mapsto Y_{g,t}(d_1,...,d_t)$ is increasing in each of its arguments. By contrast,
one can show that when $D_{g,t}=I_g1\{t\geq F\}$,
$$\delta^n_\ell=\sum_{g:I_g\ne 0}\frac{I_g}{\sum_{g':I_{g'}\ne 0}I_{g'}}\frac{\delta_{g,\ell}}{I_g\ell},$$
a weighted average of group $g$'s effect of having received $I_g$ rather than $0$ units of treatment for $\ell$ periods, normalized by $I_g\ell$, with only non-negative weights. Therefore, $\delta^n_\ell$ satisfies
the no-sign reversal property. The normalization of $\delta_{g,\ell}$ by $I_g\ell$ ensures that $\delta^n_\ell$ can be interpreted as a weighted average of the slopes of the potential outcome function with respect to the current treatment and its first $\ell-1$ lags. By contrast, the implicit normalization by $I_g$ in $E\left(\widehat{\beta}_{fe,\ell}|\bm{D}\right)$  is hard to justify: at $F_g-1+\ell$, $g$ has received $I_g\ell$ more treatment doses
than in the status-quo counterfactual. Point 2 of Proposition \ref{prop:regression1} shows that for $\ell \in \{1,...,F-2\}$, $E\left(\widehat{\beta}_{fe,\ell}|\bm{D}\right)=0$ under Assumptions \ref{hyp:no_antic} and \ref{hyp:strong_exogeneity}. Accordingly, testing $\beta_{fe,\ell}=0$ is a valid test of Assumptions \ref{hyp:no_antic} and \ref{hyp:strong_exogeneity}.

\subsection{Local-projection panel regressions.}

\paragraph{Local-projection panel regressions.}
Among the nine papers in our census that do not have a binary and staggered design and estimate dynamic effects, the second estimation method we found are regressions of leads of the outcome on group and period FEs and the treatment. Such regressions have sometimes been described as a panel-data version of the local-projection method originally proposed by \cite{jorda2005estimation} for time-series data.
\begin{reg}\label{reg:leads_fd_on_d}
(Local-projection panel regressions) For every $\ell \in \{1,...,T-1\}$ such that the regression that follows is well-defined, let $\widehat{\beta}_{lp,\ell}$ denote the coefficient on $D_{g,t}$ in a regression of $Y_{g,t-1+\ell}$ on group and period FEs and $D_{g,t}$, in the subsample such that $1\leq t\leq T-\ell+1$.
\end{reg}
Intuitively, researchers hope that $\widehat{\beta}_{lp,\ell}$ estimates some average effect of increasing groups' period-$t$ treatment by one unit on their period $t-1+\ell$ outcome. Researchers may estimate Regression \ref{reg:leads_fd_on_d} in first-difference. A result similar to that in Proposition \ref{prop:regression4} below also applies to that specification.

\paragraph{A decomposition of the coefficients in local-projection panel regressions.}
We decompose local-projection panel regressions in Design \ref{des:untreatedatbaseline}: the papers in our census that have used those regressions fall into that design. For all $\ell \in \{1,...,T-1\}$, let $\widehat{\eps}^\ell_{g,t}$ denote the residual from a regression of $D_{g,t}$ on group and time fixed effects, in the subsample such that $1\leq t\leq T-\ell+1$.
For all $g$ such that $F_g\le T$ and all $k\in \{1,...,T-F_g+1\}$, let $$\bar{D}_{g,k}=\frac{1}{k}\sum_{t=F_g}^{F_g-1+k} D_{g,t}$$ denote $g$'s average treatment from $F_g$ to $F_g-1+k$. Finally, let $\underline{F}=\min_g F_g$.
\begin{prop}\label{prop:regression4}
Suppose that we are in Design \ref{des:untreatedatbaseline}, and Assumptions \ref{hyp:no_antic}-\ref{hyp:strong_exogeneity} hold.
\begin{enumerate}
\item $\forall \ell \in \{1,...,T-1\}$ such that $\widehat{\beta}_{lp,\ell}$ is well-defined,
\begin{equation*}
E[\widehat{\beta}_{lp,\ell}|\bm{D}]= \sum_{k=1}^{T-\underline{F}+1} \sum_{g:\ell-k+1 \le F_g \le T-k+1}  w^{lp,\ell}_{g,k} \frac{\delta_{g,k}}{\bar{D}_{g,k}},
\end{equation*}
where
$$w_{g,k}^{lp,\ell} = \frac{\bar{D}_{g,k}\widehat{\eps}^\ell_{g,F_g-\ell+k}}{\sum_{k'=\ell}^{T-\underline{F}+1}\sum_{g':F_{g'}\le T-k'+1}  D_{g',F_{g'}-\ell+k'}\widehat{\eps}^\ell_{g',F_{g'}-\ell+k'}}.$$
\item In Design \ref{des:binaryandstaggered}, $\min_{g,k} w_{g,k}^{lp,\ell}<0$ for all $\ell \ge 2$ and for all $\ell\in\{2,...,\underline{F}\}$, $$\sum_{k=1}^{T-\underline{F}+1} \sum_{g:\ell-k+1\le F_g-1+k\le T}  w^{lp,\ell}_{g,k}< 1.$$
\item In Design \ref{des:stag_nonbinary}, $\sum_{k=1}^{T-\underline{F} +1}\sum_{g:F_g-1+k\le T}  w^{lp,1}_{g,k}= 1$.
\end{enumerate}
\end{prop}

\paragraph{Interpretation of, and some remarks on, Proposition \ref{prop:regression4}.}
Proposition \ref{prop:regression4} shows that under Design \ref{des:untreatedatbaseline}, $\widehat{\beta}_{lp,\ell}$ estimates a weighted sum of $\delta_{g,k}$, normalized by $\bar{D}_{g,k}$. The weighted sum is across groups, and across the number $k$ of periods of exposure to higher treatment doses. Accordingly, $\widehat{\beta}_{lp,\ell}$ does not estimate an average across groups of the effect of $\ell$ periods of exposure to higher treatment doses: $\widehat{\beta}_{lp,\ell}$ is contaminated by effects of other lengths of exposure. It is easy to understand where this contamination issue stems from in binary and staggered designs. Then, some groups with $D_{g,t}=1$ may have started receiving the treatment before period $t$, so the local-projection regression of $Y_{g,t-1+\ell}$ on $D_{g,t}$ captures an effect of more than $\ell$ periods of exposure for those groups. Similarly, some groups with $D_{g,t}=0$, which are supposed to be control groups, may have started receiving treatment between periods $t+1$ and $t-1+\ell$, and the local-projection regression of $Y_{g,t-1+\ell}$ on $D_{g,t}$ captures an effect of less than $\ell$ periods of exposure for those groups. Beyond this contamination phenomenon, a further issue is that some of the weights may be negative, and we actually show that for $\ell \geq 2$, some weights are always negative in Design \ref{des:binaryandstaggered}. A last and perhaps even more concerning issue is that in Design \ref{des:binaryandstaggered}, for $\ell\in\{2,...,\underline{F}\}$, the weights $w_{g,t}^{lp,\ell}$ sum to strictly less than one. This implies that even if there is a, say, positive real number $\theta$ such that $\delta_{g,k}/\bar{D}_{g,k}=\theta$, meaning that the treatment effect does not vary with length of exposure $k$ or across groups $g$, $E\left(\widehat{\beta}_{lp,\ell}\right)< \theta$: the local-projection regression is downward biased. This is because the regression is misspecified: it considers groups with $D_{g,t}=0$ as untreated, whereas some of them may actually be treated at $t-1+\ell$.

\paragraph{Connection with other literature on local projection regressions.}
The issues with local-projection regressions we highlight in Proposition \ref{prop:regression4} are specific to Regression \ref{reg:leads_fd_on_d}, an extension of local projection to panel data. Those issues are absent with time-series data where the unit receives a single shock, the design considered by \cite{jorda2005estimation}. Those issues are also absent in the careful extension of local projection regressions to panel data with binary and staggered treatments recently proposed by \cite{dube2023local}, in work posterior to ours.

\subsection{Distributed-lag regressions.}

\paragraph{Distributed-lag regressions.} Among the nine papers in our census that do not have a binary and staggered design and estimate dynamic effects, the third estimation method we found are distributed-lag regressions.
\begin{reg}\label{reg:twfe_lags}
(Distributed-lag regression) For $l \in \{0,...,K\}$, let $\widehat{\beta}_{dl,l}$ denote the coefficient on $D_{g,t-l}$ in a regression of $Y_{g,t}$ on group and period FEs and $\left(D_{g,t-l}\right)_{l \in \{0,...,K\}}$, in the subsample such that $t\geq K+1$.
\end{reg}
Regression \ref{reg:twfe_lags} is discussed in \cite{angrist2008mostly}, see Equation (5.2.6) therein. In practice, researchers may slightly augment or modify Regression \ref{reg:twfe_lags}. They may include treatment leads in the regression, to test for parallel trends. They may define the lagged treatments as equal to 0 at time periods when they are not observed, and estimate the regression in the full sample. They may also estimate the regression in first difference and without group fixed effects. Results similar to Proposition \ref{prop:regression3} below apply to all those variations on Regression \ref{reg:twfe_lags}.

\paragraph{A decomposition of the coefficients in distributed-lag regressions.}
\begin{prop}\label{prop:regression3}
(Application of Corollary 1 in \cite{dCDH_several} to Regression \ref{reg:twfe_lags})
If Regression \ref{reg:twfe_lags} is well-defined, Assumptions \ref{hyp:no_antic}-\ref{hyp:strong_exogeneity} hold  and for all $g$ and $t\geq K+1$ there exists real numbers $\left(\gamma^l_{g,t}\right)_{l \in \{0,...,K\}}$ such that for all $\bm{d} \in \{0,1\}^t$, $Y_{g,t}(\bm{d})=Y_{g,t}(\bm{0}_t)+\sum_{l=0}^{K}\gamma^l_{g,t}d_{t-l}$, then for all $l \in \{0,...,K\}$,
$$E\left[\widehat{\beta}_{dl,l}|\bm{D}\right]=\sum_{\substack{(g,t):D_{g,t-l}\ne 0, \\ t\geq K+1}}w^{dl,l,l}_{g,t}\gamma^{l}_{g,t}+\sum_{\substack{l'=0 \\ l'\neq l}}^{K}~\sum_{\substack{(g,t):D_{g,t-l'}\ne 0, \\ t\geq K+1}}w^{dl,l,l'}_{g,t}\gamma^{l'}_{g,t},$$
where, for $l=l'$ or $l\ne l'$, the weights $w^{dl,l,l'}_{g,t}$ are defined by
$$w^{dl,l,l'}_{g,t}:=\frac{D_{g,t-l'}\,\eta_{g,t}^l}{\sum_{(g',t'):t'\ge K+1}D_{g',t'-l}\,\eta_{g',t'}^l},$$
with $(\eta_{g,t}^l)_{g,t}$ the residuals in a regression of $D_{g,t-l}$ on period and group fixed effects and $\left(D_{g,t-l'}\right)_{l'\in \{0,...,K\},l'\neq l}$. Moreover,
$$\sum_{\substack{(g,t):D_{g,t-l}\ne 0, \\ t\geq K+1}}w^{dl,l,l}_{g,t}=1, \quad \sum_{\substack{(g,t):D_{g,t-l'}\ne 0, \\ t\geq K+1}}w^{dl,l,l'}_{g,t}=0 \quad \forall l' \ne l.$$
\end{prop}

\paragraph{Interpretation of, and some remarks on, Proposition \ref{prop:regression3}.} Proposition \ref{prop:regression3} is an application of Corollary 1 in \cite{dCDH_several}, a more general result applicable to any TWFE regression with several treatments, to distributed-lag regressions. Corollary 1 in \cite{dCDH_several} was only proven for binary treatments, but extending it to non-binary treatments is straightforward. Proposition \ref{prop:regression3} is also related to the results of \cite{abraham2018} for event-study regressions in binary-and-staggered designs. To obtain this decomposition, we assume that $$Y_{g,t}(\bm{d})=Y_{g,t}(\bm{0}_t)+\sum_{l=0}^{K}\gamma^l_{g,t}d_{t-l},$$
meaning that the functional form of the distributed-lag regression is correctly specified: only the first $K$ treatment lags affect the outcome, and those lags do not interact. Even under those strong assumptions, Proposition \ref{prop:regression3} shows that $\widehat{\beta}_{dl,l}$, the coefficient on the $l$th treatment lag, may not estimate a well-defined causal effect. Specifically, Proposition \ref{prop:regression3} shows that this coefficient estimates the sum of $K+1$ terms. The first term is a weighted sum of the effect of the $l$th treatment lag, across all $(g,t)$ cells for which that lag is not equal to $0$, with weights that sum to one but may be negative. This term may be biased for the average effect of the $l$th treatment lag, if that effect varies across $(g,t)$ cells. The remaining $K$ terms are weighted sums of the effects of other treatment lags, with weights summing to zero. If the effects of the other lags vary across $(g,t)$ cells, those terms may differ from zero and may contaminate $\widehat{\beta}_{dl,l}$.\footnote{\label{footnote:twowayfeweights} The \st{twowayfeweights} command can be used to compute the weights attached to distributed-lag regressions, using the \st{other\_treatments} option.}

\section{Application to banking deregulations and the housing market}\label{sec:appli}

\subsection{Setting and research question}

In 1994, the Interstate Banking and Branching Efficiency Act (IBBEA) allowed US Banks to operate across state borders without formal authorization from state authorities. Initially, all states still imposed restrictions on: de novo branching without explicit agreement by state authorities; the minimum age of the target institution in case of mergers; the acquisition
of individual branches without acquiring the entire bank; the total amount
of statewide deposits controlled by a single bank or bank holding company.
\cite{rice2010does} compute the number of restrictions in place in each state and year from 1994 to 2005. \cite{favara2015credit} reverse their index, so their treatment is the number of restrictions lifted, ranging from 0 to 4. Several papers in the finance literature have aggregated these qualitatively different deregulations into a scalar variable. As the treatment is equal to 0 in every state in 1994, our event-study parameters are just average effects of having experienced some deregulations versus none, and do not rely on any homogeneity assumption on the effects of different policies.\footnote{Since no state had conducted deregulations prior to 1994, the initial-conditions problem is also not an issue.} \cite{favara2015credit} use 1994-to-2005 county-level data to estimate the effect of the number of regulations lifted on the growth of mortgages originated by banks, and on the growth of houses prices.

\subsection{Local-projection-regressions}

\paragraph{Regression model.}
\cite{favara2015credit} use a panel data version of the local-projection regressions proposed by \cite{jorda2005estimation}. Let $\Delta$ denote the first-difference operator. To estimate the treatment effect on, say, the growth rate of loan volume, they regress, for every $\ell \in \{1,...,9\}$, $\Delta \ln(L_{g,t-1+\ell})$, the log growth rate of loans in county $g$ in year $t-1+\ell$, on county and year FEs, $D_{g,t-1}$, the number of deregulations in county $g$ and year $t-1$, and some controls. Proposition \ref{prop:regression4} does not apply to their specification, because the authors define their treatment variable as $D_{g,t-1}$, the lagged treatment, and because the regression has some controls. Thus, we reestimate their local-projection regression, with $D_{g,t}$ instead of $D_{g,t-1}$ and without controls. The top-left panel of Figure \ref{figure_FavaraImbs} below shows that doing so, we find significant and positive coefficients $\widehat{\beta}_{lp,\ell}$ till $\ell=5$, insignificant coefficients at $\ell=6$, $7$, and $9$, and a significantly negative coefficient at $\ell=8$. This is in line with the results the authors obtained with their regression, except that with their specification the coefficient remains positive and significant till $\ell=4$ and is not significantly negative at $\ell=8$ (see their Figure 1, Panel B). These results lead the authors to conclude that the growth effects of deregulation on credit supply are temporary. The bottom-left panel of Figure \ref{figure_FavaraImbs} shows that results are similar for the log growth rate of houses prices: coefficients are marginally significant and positive till $\ell=4$, and insignificant after.

\paragraph{Decomposition of the local-projection coefficients.} We use Proposition \ref{prop:regression4} to decompose the local-projection coefficients shown on the top-left panel of Figure \ref{figure_FavaraImbs}.\footnote{Results are very similar though not completely identical for the local-projection coefficients shown on the bottom-left panel, because counties with missing values are not exactly the same for the two outcomes.} $\widehat{\beta}_{lp,1}$ is a weighted sum of 7,626 effects $\delta_{g,k}/\bar{D}_{g,k}$, where 4,670 effects are weighted positively and 2,956 effects are weighted negatively, and where positive and negative weights respectively sum to $1.067$ and $-0.125$. Weights do not sum to one, because this application does not meet the conditions in Design \ref{des:stag_nonbinary}, but their sum is still not very far from one. $\widehat{\beta}_{lp,1}$, which is supposed to measure the effect of one year of exposure to treatment, is contaminated by effects of other lengths of exposure: weights on effects of one year of exposure sum to $0.294$, and weights on effects of other lengths of exposure sum to $0.648$.
$\widehat{\beta}_{lp,2}$ is a weighted sum of 7,626 effects, where 4,424 effects are weighted positively and 3,202 effects are weighted negatively, and where positive and negative weights sum to $1.085$ and $-0.584$. As in Point 3 of Proposition \ref{prop:regression4}, and even though the design is not binary and staggered, the weights sum to strictly less than one, and they actually sum to much less than one. Then, even if $\delta_{g,k}/\bar{D}_{g,k}$ did not vary across $g$ or $k$, $\widehat{\beta}_{lp,2}$ would be severely biased towards zero. $\widehat{\beta}_{lp,2}$ is contaminated by effects of other lengths of exposure than two years. Most of those effects are weighted positively, except for the effects of one year of exposure, whose weights sum to $-0.472$. Intuitively, this is due to the fact that groups with $D_{g,t}=0,D_{g,t+1}>0$ are used as ``control groups'' by $\widehat{\beta}_{lp,2}$, whereas they are treated at $t+1$. Results are similar for $\widehat{\beta}_{lp,3}$, except that weights now only sum to $0.069$, and effects of one and two years of exposure are weighted negatively, with weights summing to $-0.429$ and $-0.478$, respectively. For $\widehat{\beta}_{lp,4}$, weights sum to $-0.018$. This implies that even if $\delta_{g,k}/\bar{D}_{g,k}$ did not vary across $g$ or $k$, $E\left[\widehat{\beta}_{lp,4}\right]$ would be of a different sign than the treatment effect. Results for $\widehat{\beta}_{lp,5}$ to $\widehat{\beta}_{lp,9}$ are similar to the results for $\widehat{\beta}_{lp,4}$. In particular, the sum of weights is negative for those coefficients.

\subsection{$\DID_\ell$ and $\DID^n_\ell$ estimators.}

\paragraph{Details on the design.}

Eight states never deregulate. 33 states deregulate only once, eight states deregulate twice, and one state deregulates three times. Of these 42 states that deregulate at least once, 38 do so for the first time in 1995, 1996, 1997, or 1998. Accordingly, up to $\ell=2005-1998+1=8$, the $\delta_{\ell}$ parameters apply to similar sets of counties: $\delta_{1}$ and $\delta_{8}$ respectively apply to 905 and 773 counties. $\delta_{9}$, $\delta_{10}$, and $\delta_{11}$ on the other hand only apply to a smaller subsample of counties (357, 238, and 1 county, respectively). Hence, we do not report
estimates of those parameters.\footnote{For the two outcomes we consider, the
estimates of $\delta_{9}$ and $\delta_{10}$ are even larger than those of $\delta_{8}$.} Similarly, five placebo estimators can be computed, but only three apply to more than 50\% of the 905 counties whose treatment changes at least once. The four and fifth placebos only apply to 128 and 120 counties respectively, so we do not report them.\footnote{Numbers are for the volume-of-mortgages-originated outcome; numbers would slightly differ for the houses-price outcome: counties with missing values are not exactly the same for the two outcomes.}

\begin{figure}[H]
\begin{center}
\includegraphics[trim=7mm 5mm 5mm 2mm, clip=true, scale=1.2]{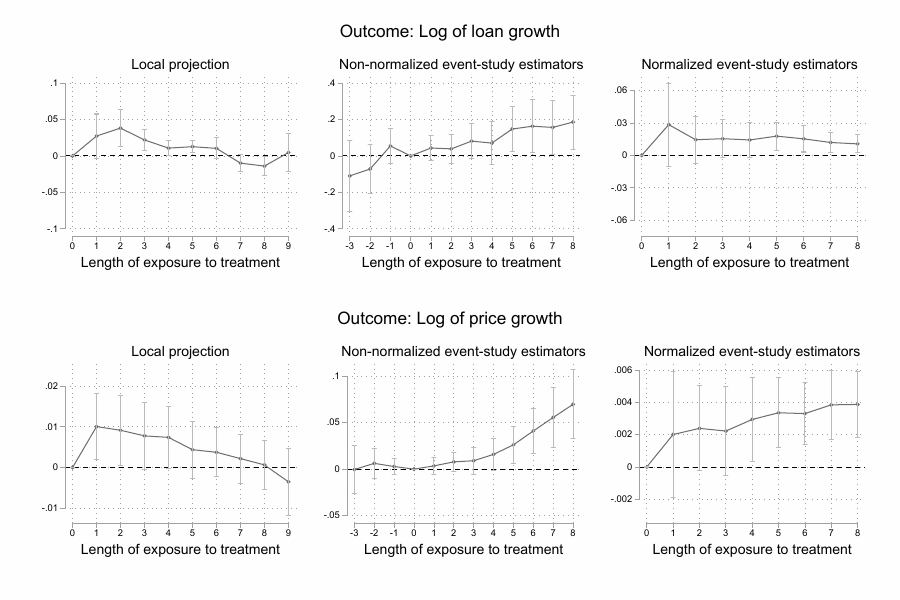}
\caption{Effect of banking deregulations on loan volume and houses prices.}
\label{figure_FavaraImbs}
\end{center}
\begin{minipage}[c]{1\textwidth}
 {\footnotesize{Notes. The top-left panel shows estimated effects of banking deregulations on loan volume, according to local-projection regressions of $\Delta \ln(L_{g,t-1+\ell})$, the log growth rate of loans in county $g$ in year $t-1+\ell$, on county and year FEs, and $D_{g,t}$, the number of deregulations in county $g$ and year $t$, for $\ell \in \{1,...,9\}$.  To the right of zero, the top-centre panel shows $\DID_{\ell}$ estimates of the effect of having been exposed to banking deregulations for $\ell$ periods on the log growth rate of loans, for $\ell \in \{1,...,8\}$. To the left of zero, the top-center panel shows $\DID^{\pl}_{\ell}$ placebo estimates, for $\ell \in \{1,...,3\}$. The top-right panel shows the $\DID^n_{\ell}$ estimates, for $\ell \in \{1,...,8\}$. The bottom-left, bottom-centre, and bottom-right panels are similar to the top-left, top-centre, and top-right panels, with houses prices instead of loans as the outcome. Estimates in the bottom panels are weighted by the inverse of the number of counties per state: \cite{favara2015credit} weight their regressions by that variable when they study the effect of deregulations on houses prices. 95\% confidence intervals relying on a normal approximation and standard errors clustered at the state level are shown in red on all panels. All estimates are computed using the 1994-2005 county-level panel data set constructed by \cite{favara2015credit}.}}
\end{minipage}
\end{figure}

\paragraph{$\DID_\ell$ estimators, for the loan-volume outcome.}
The top-center panel of Figure \ref{figure_FavaraImbs} plots the $\DID_\ell$ estimators for the loan-volume outcome to the right of zero, and the corresponding placebo estimators to the left. $\DID_{1}=0.043$ (s.e.=$0.035$): after one year of exposure to deregulation, the loan-volume growth rate increases by 4.3 percentage points more in counties that deregulate than in counties that do not, an insignificant difference. This effect builds up over time, and becomes significant at the 10\% level after three years of exposure  ($\DID_{3}=0.081$, s.e.=$0.049$), and at the 5\% level after five years of exposure ($\DID_{5}=0.148$, s.e.=$0.064$). To the left of zero, placebo
estimates are shown. Placebos are jointly insignificant (F-test p-value=$0.400$). But the placebos are relatively large, and imprecisely estimated. For instance, several $\DID_{\ell}$ estimates are below the line connecting the lower bound of the confidence interval of $\DID^{\pl}_{1}$ and $(0,0)$. Moreover, we can only test the parallel-trends assumption over three years, whereas that assumption has to hold for eight years for $\DID_{8}$ to be unbiased. Overall, even if the placebos do not clearly indicate a violation of the parallel-trends assumption, they may fail to detect violations of parallel trends large enough to substantially bias the $\DID_{\ell}$ estimates \citep[see][]{roth2019pre}.

\paragraph{$\DID^n_\ell$ estimators, for the loan-volume outcome.}
The top-right panel of Figure \ref{figure_FavaraImbs} plots the $\DID^n_\ell$ estimators for the loan-volume outcome. In order to test for Point 1 of Lemma \ref{lem:norm_test}, we restrict the sample of switchers to the 773 counties for which $\DID_{g,8}$ can be computed, and further discard 141 switchers in states that experienced more than one treatment change (results are not very different if one does not restrict the sample). $\DID^n_\ell$ are estimators of the effect of the current treatment and of its $\ell-1$ first lags on the outcome. In this restricted sample, each lag receives a weight equal to $1/\ell$. With the exception of $\DID^n_1$, which is much larger than the other estimates but also imprecisely estimated, all estimates are relatively close to each other, suggesting that treatment effects are relatively stable over time, and that the effects of the current and lagged treatments are not very different. To formally test this, we follow Point 1 of Lemma \ref{lem:norm_test} in the Web Appendix, and test whether $\ell\mapsto \DID^n_\ell$ is constant. The test is not rejected (p-value = 0.328), which goes against the authors' conclusion that deregulations only have short-lived effects on mortgage volume. Our decompositions of their local projection estimators suggest that their coefficients may become small and insignificant because the sum of the weights attached to $\widehat{\beta}_{lp,\ell}$ decreases as $\ell$ increases.

\paragraph{$\DID_\ell$ and $\DID^n_\ell$ estimators, for the houses-prices outcome.}
The bottom-center panel of Figure \ref{figure_FavaraImbs}  plots the $\DID_\ell$ estimators for the houses-prices outcome. $\DID_{1}=0.003$ (s.e.=$0.004$): after one year of exposure to deregulation, the houses-price growth rate increases by 0.3 percentage points more in counties that deregulate than in counties that do not, an insignificant difference. This effect builds up over time, and becomes significant at the 10\% level after four years of exposure  ($\DID_{4}=0.016$, s.e.=$0.009$), and at the 5\% level after five years of exposure ($\DID_{5}=0.026$, s.e.=$0.010$). Placebos are jointly insignificant (F-test p-value=$0.703$). For this outcome, the placebos are small, fairly precisely estimated, and suggest that if anything, differential pre-trends would downward-bias the estimated effects. The bottom-right panel of Figure \ref{figure_FavaraImbs}  plots the $\DID^n_\ell$ estimators, restricting the sample of switchers to counties for which $\DID_{g,8}$ can be computed and in states that experienced only one treatment change. Estimates increase with $\ell$:
deregulations seem to also have long-lasting effects on houses prices.

\newpage

\linespread{1.2}\selectfont
\bibliographystyle{chicago}
\bibliography{biblio}

@unpublished{small2007stochastic,
  title={A stochastic monotonicity assumption for the instrumental variables method},
  author={Small, Dylan and Tan, Zhiqiang},
  note={Working paper},
  year={2007}
}

@article{burgess2015value,
  title={The value of democracy: evidence from road building in Kenya},
  author={Burgess, Robin and Jedwab, Remi and Miguel, Edward and Morjaria, Ameet and Padr{\'o} i Miquel, Gerard},
  journal={American Economic Review},
  volume={105},
  number={6},
  pages={1817--51},
  year={2015}
}

@article{pierce2016surprisingly,
  title={The surprisingly swift decline of US manufacturing employment},
  author={Pierce, Justin R and Schott, Peter K},
  journal={American Economic Review},
  volume={106},
  number={7},
  pages={1632--62},
  year={2016}
}

@article{fuest2018higher,
  title={Do higher corporate taxes reduce wages? Micro evidence from Germany},
  author={Fuest, Clemens and Peichl, Andreas and Siegloch, Sebastian},
  journal={American Economic Review},
  volume={108},
  number={2},
  pages={393--418},
  year={2018}
}

@unpublished{callaway2021difference,
  title={Difference-in-differences with a continuous treatment},
  author={Callaway, Brantly and Goodman-Bacon, Andrew and Sant'Anna, Pedro HC},
  note={arXiv preprint arXiv:2107.02637},
  year={2021}
}

@article{krolikowski2018choosing,
  title={Choosing a control group for displaced workers},
  author={Krolikowski, Pawel},
  journal={ILR Review},
  volume={71},
  number={5},
  pages={1232--1254},
  year={2018},
  publisher={SAGE Publications Sage CA: Los Angeles, CA}
}

@unpublished{de2022difference,
  title={Difference-in-Differences for Continuous Treatments and Instruments with Stayers},
  author={de Chaisemartin, Cl{\'e}ment and D'Haultf{\oe}uille, Xavier and Pasquier, F{\'e}lix and Vazquez-Bare, Gonzalo},
  note={arXiv preprint arXiv:2201.06898},
  year={2022}
}

@article{de2022survey,
  title={Two-way fixed effects and differences-in-differences with heterogeneous treatment effects: A survey},
  author={de Chaisemartin, Cl{\'e}ment and D'Haultf{\oe}uille, Xavier},
  journal={The Econometrics Journal},
  volume={26},
  number={3},
  pages={C1--C30},
  year={2023},
  publisher={Oxford University Press}
}

@article{Murphy2001,
  title={Marginal mean models for dynamic regimes},
  author={Murphy, Susan A and van der Laan, Mark J and Robins, James M and Conduct Problems Prevention Research Group},
  journal={Journal of the American Statistical Association},
  volume={96},
  number={456},
  pages={1410--1423},
  year={2001},
  publisher={Taylor \& Francis}
}

@unpublished{chaisemartin2020intertemporal,
    title={Difference-in-Differences Estimators of Intertemporal Treatment Effects},
    author={Clément de Chaisemartin and Xavier D'Haultfœuille},
    year={2020},
    note={arXiv preprint 2007.04267v8},
    archivePrefix={arXiv},
    primaryClass={econ.EM}
}

@article{roth2019pre,
  title={Pretest with caution: Event-study estimates after testing for parallel trends},
  author={Roth, Jonathan},
  journal={American Economic Review: Insights},
  volume={4},
  number={3},
  pages={305--22},
  year={2022}
}

@inproceedings{beel2009google,
  title={Google Scholar's ranking algorithm: an introductory overview},
  author={Beel, J{\"o}ran and Gipp, Bela},
  booktitle={Proceedings of the 12th international conference on scientometrics and informetrics (ISSI'09)},
  volume={1},
  pages={230--241},
  year={2009},
  organization={Rio de Janeiro (Brazil)}
}

@article{deryugina2017fiscal,
  title={The fiscal cost of hurricanes: Disaster aid versus social insurance},
  author={Deryugina, Tatyana},
  journal={American Economic Journal: Economic Policy},
  volume={9},
  number={3},
  pages={168--98},
  year={2017}
}

@article{jorda2005estimation,
  title={Estimation and inference of impulse responses by local projections},
  author={Jord{\`a}, {\`O}scar},
  journal={American Economic Review},
  volume={95},
  number={1},
  pages={161--182},
  year={2005}
}

@article{favara2015credit,
  title={Credit supply and the price of housing},
  author={Favara, Giovanni and Imbs, Jean},
  journal={American Economic Review},
  volume={105},
  number={3},
  pages={958--92},
  year={2015}
}

@article{rice2010does,
  title={Does credit competition affect small-firm finance?},
  author={Rice, Tara and Strahan, Philip E},
  journal={The Journal of Finance},
  volume={65},
  number={3},
  pages={861--889},
  year={2010},
  publisher={Wiley Online Library}
}

@article{bojinov2020panel,
  title={Panel experiments and dynamic causal effects: A finite population perspective},
  author={Bojinov, Iavor and Rambachan, Ashesh and Shephard, Neil},
  journal={Quantitative Economics},
  volume={12},
  number={4},
  pages={1171--1196},
  year={2021},
  publisher={Wiley Online Library}
}

@article{robins1986new,
  title={A new approach to causal inference in mortality studies with a sustained exposure period-application to control of the healthy worker survivor effect},
  author={Robins, James},
  journal={Mathematical modelling},
  volume={7},
  number={9-12},
  pages={1393--1512},
  year={1986},
  publisher={Elsevier}
}

@article{malani2015,
  title={Interpreting pre-trends as anticipation: Impact on estimated treatment effects from tort reform},
  author={Malani, Anup and Reif, Julian},
  journal={Journal of Public Economics},
  volume={124},
  pages={1--17},
  year={2015},
  publisher={Elsevier}
}

@article{botosaru2018difference,
  title={Difference-in-differences when the treatment status is observed in only one period},
  author={Botosaru, Irene and Gutierrez, Federico H},
  journal={Journal of Applied Econometrics},
  volume={33},
  number={1},
  pages={73--90},
  year={2018},
  publisher={Wiley Online Library}
}

@article{east2023multi,
Author = {East, Chloe N. and Miller, Sarah and Page, Marianne and Wherry, Laura R.},
Title = {Multigenerational Impacts of Childhood Access to the Safety Net: Early Life Exposure to Medicaid and the Next Generation's Health},
Journal = {American Economic Review},
Volume = {113},
Number = {1},
Year = {2023},
Pages = {98-135}
}

@unpublished{dube2023local,
  title={A local projections approach to difference-in-differences event studies},
  author={Dube, Arindrajit and Girardi, Daniele and Jorda, Oscar and Taylor, Alan M},
  year={2023},
  note={National Bureau of Economic Research WP 31184}
}

@article{dCDH_several,
  title={Two-way fixed effects and differences-in-differences estimators with several treatments},
  author={de Chaisemartin, Cl{\'e}ment and D'Haultf{\oe}uille, Xavier},
  journal={Journal of Econometrics},
  volume={236},
  number={2},
  pages={105480},
  year={2023},
  publisher={Elsevier}
}

@article{abbring2003nonparametric,
  title={The nonparametric identification of treatment effects in duration models},
  author={Abbring, Jaap H and Van den Berg, Gerard J},
  journal={Econometrica},
  volume={71},
  number={5},
  pages={1491--1517},
  year={2003},
  publisher={Wiley Online Library}
}

@article{borusyak2020revisiting,
  title={Revisiting event study designs: Robust and efficient estimation},
  author={Borusyak, Kirill and Jaravel, Xavier and Spiess, Jann},
  journal={Review of Economic Studies},
  volume={91},
  year={2024},
  pages={3253--3285}
}

@unpublished{harmon2022difference,
  title={Difference-in-Differences and Efficient Estimation of Treatment Effects},
  author={Harmon, Nikolaj A},
  note={working paper},
  year={2022}
}

@book{angrist2008mostly,
  title={Mostly harmless econometrics},
  author={Angrist, Joshua D and Pischke, J{\"o}rn-Steffen},
  year={2008},
  publisher={Princeton university press}
}

@article{abraham2018,
  title={Estimating dynamic treatment effects in event studies with heterogeneous treatment effects},
  author={Sun, Liyang and Abraham, Sarah},
  journal={Journal of Econometrics},
  volume={225},
  number={2},
  pages={175--199},
  year={2021},
  publisher={Elsevier}
}

@unpublished{de2023did_multiplegt_dyn,
  title={DID\_MULTIPLEGT\_DYN: Stata module to estimate event-study Difference-in-Difference (DID) estimators in designs with multiple groups and periods, with a potentially non-binary treatment that may increase or decrease multiple times},
  author={de Chaisemartin, Cl{\'e}ment and D'Haultfoeuille, Xavier and Mal{\'e}zieux, M{\'e}litine and Sow, Doulo},
  year={2023},
  note={Boston College Department of Economics}
}

@article{callaway2018,
  title={Difference-in-differences with multiple time periods},
  author={Callaway, Brantly and Sant’Anna, Pedro HC},
  journal={Journal of Econometrics},
  volume={225},
  number={2},
  pages={200--230},
  year={2021},
  publisher={Elsevier}
}

@article{goodman2018,
  title={Difference-in-differences with variation in treatment timing},
  author={Goodman-Bacon, Andrew},
  journal={Journal of Econometrics},
  volume={225},
  number={2},
  pages={254--277},
  year={2021},
  publisher={Elsevier}
}

@article{bertrand2004,
  title={How much should we trust differences-in-differences estimates?},
  author={Bertrand, Marianne and Duflo, Esther and Mullainathan, Sendhil},
  journal={The Quarterly Journal of Economics},
  volume={119},
  number={1},
  pages={249--275},
  year={2004},
  publisher={MIT Press}
}

@article{dCdH_AER,
  title={Two-way fixed effects estimators with heterogeneous treatment effects},
  author={de Chaisemartin, Clement and D'Haultf{\oe}uille, Xavier},
   journal = {American Economic Review},
  year={2020},
  volume={110},
	number={9},
	pages={2964-2996}
}

@unpublished{borusyak2016,
  title={Revisiting event study designs},
  author={Borusyak, Kirill and Jaravel, Xavier},
  note={Working Paper},
  year={2017}
}

@Article{Abadie05,
  author={Alberto Abadie},
  title={Semiparametric Difference-in-Differences Estimators},
  journal={Review of Economic Studies},
  year={2005},
  volume={72},
  number={1},
  pages={1-19},
  month={01},
}

@article{deChaisemartin15b,
author = {Cl\'ement de Chaisemartin and  Xavier D'Haultf{\oe}uille},
title = {Fuzzy Differences-in-Differences},
journal = {The Review of Economic Studies},
volume = {85},
number = {2},
pages = {999--1028},
year = {2018}
}

\storecounter{equation}{nbeq}
\storecounter{lem}{nblem}
\storecounter{hyp}{nbhyp}
\storecounter{thm}{nbthm}
\storecounter{conddes}{nbconddes}

\newpage
\includepdf[pages=-]{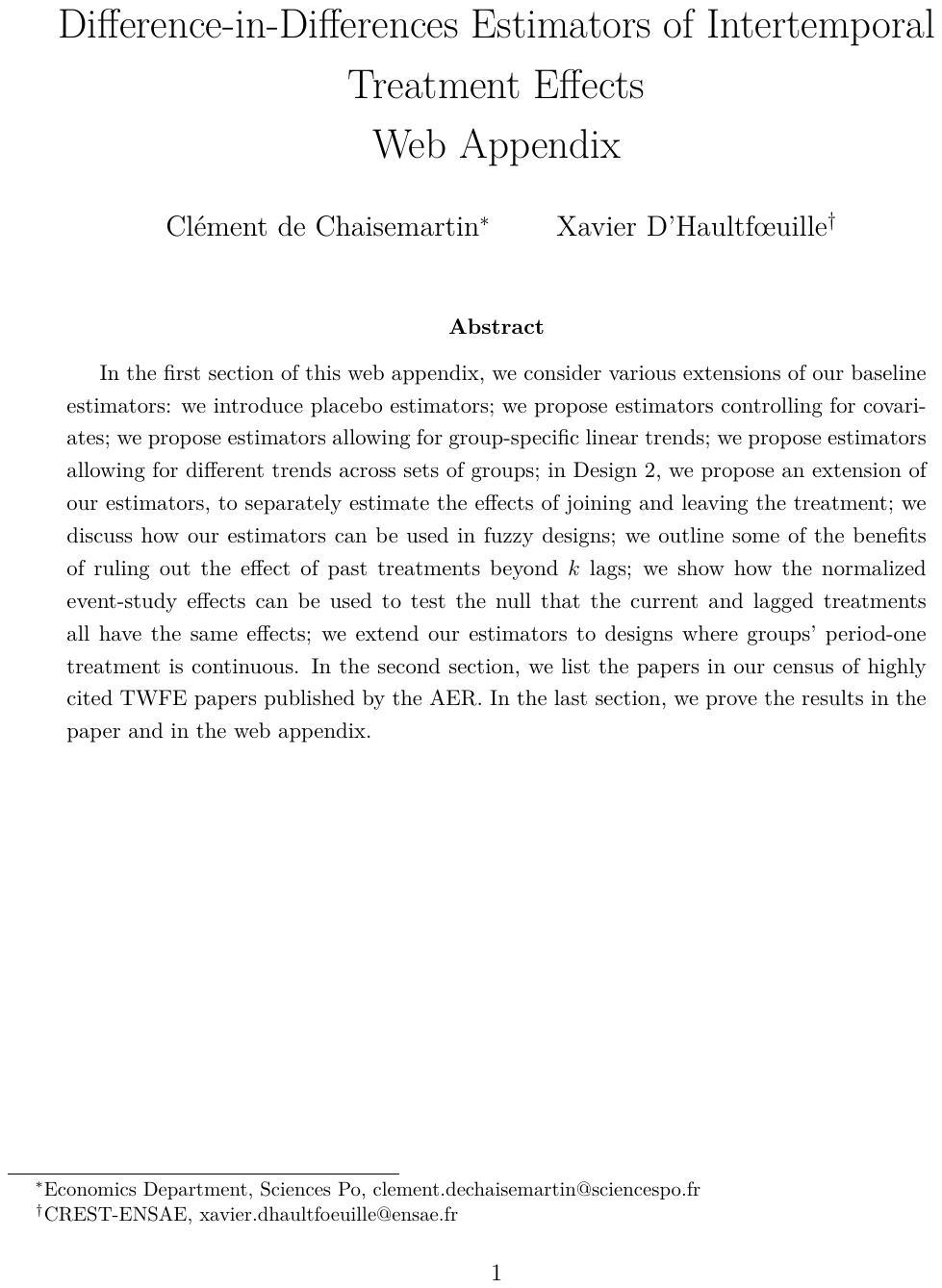}

\end{document}